\documentclass{./emulateapj}

\usepackage{lscape}

\begin{document}


\def\mbh{{$\mathcal M_{\rm BH}$}}
\def\msol{{$\mathcal M_{\odot}$}}
\def\mr{{$M_R$}}
\def\lr{{$L_R$}}
\def\ml{{${\mathcal M}/L$}}
\def\lbulge{{$L_{\rm bulge}$}}
\def\msig{{$\mathcal M_{\rm BH}$-$\sigma_*$}}
\def\mstar{{$\mathcal M_*$}}

\shortauthors{PENG ET AL.}
\shorttitle{BLACK HOLE AND BULGE CO-EVOLUTION}

\title{PROBING THE COEVOLUTION OF SUPERMASSIVE BLACK HOLES AND
GALAXIES USING GRAVITATIONALLY LENSED QUASAR HOSTS \altaffilmark{1}}

\author {Chien Y. Peng\altaffilmark{2,3,4}, Chris D.  Impey\altaffilmark{4},
Hans-Walter Rix\altaffilmark{5}, Christopher S. Kochanek\altaffilmark{6},
Charles R Keeton\altaffilmark{7,8}, Emilio E. Falco \altaffilmark{7}, Joseph
Leh\'ar\altaffilmark{7,9}, \& Brian A. McLeod\altaffilmark{7}}

\altaffiltext{1} {Based on observations with the NASA/ESA {\it Hubble Space
Telescope}, obtained at the Space Telescope Science Institute, which is
operated by AURA, Inc., under NASA contract NAS5-26555.}

\altaffiltext{3}{STScI Fellow}

\altaffiltext{3}{Current address: Space Telescope Science Institute,
3700 San Martin Drive, Baltimore, MD 21218; cyp@stsci.edu.}

\altaffiltext{4}{Steward Observatory, University of Arizona, 933 N. Cherry
Av., Tucson, AZ 85721;  cimpey@as.arizona.edu.}

\altaffiltext{5}{Max-Planck-Institut f\"{u}r Astronomie, K\"onigstuhl 17,
Heidelberg, D-69117, Germany; rix@mpia.de}

\altaffiltext{6}{The Ohio State University, 4055 McPherson Lab, 140 West 18th
Avenue, Columbus, OH 43210}

\altaffiltext{7} {Harvard-Smithsonian Center for Astrophysics, 60 Garden
Street, Cambridge, MA 02138; falco, jlehar, bmcleod@cfa.harvard.edu}

\altaffiltext{8}{Current address: Department of Physics \& Astronomy, Rutgers
University, 136 Frelinghuysen Road, Piscataway, NJ 08854}

\altaffiltext{9} {Current address: CombinatoRx, Inc., 650 Albany Street,
Boston, MA 02118; lehar@rcn.com}

\begin {abstract}

In the present-day universe, supermassive black hole masses (\mbh) appear to
be strongly correlated with their galaxy's bulge luminosity, among other
properties.  In this study, we explore the analogous relationship between
\mbh, derived using the virial method, and the stellar $R$-band bulge
luminosity (\lr) or stellar bulge mass (\mstar) at epochs of $1 \lesssim z
\lesssim 4.5$ using a sample of 31 gravitationally lensed AGNs and 20
non-lensed AGNs.  At redshifts $z>1.7$ (10--12 Gyrs ago), we find that the
observed \mbh--\lr\ relation is nearly the same (to within $\sim 0.3$~mag) as
it is today.  When the observed \lr\ are corrected for luminosity evolution,
this means that the black holes grew in mass faster than their hosts, with the
\mbh/\mstar\ mass ratio being a factor of $\gtrsim 4^{+2}_{-1}$ times larger
at $z > 1.7$ than it is today.  By the redshift range $1 \lesssim z \lesssim
1.7$ (8--10 Gyrs ago), the \mbh/\mstar\ ratio is at most two times higher than
today, but it may be consistent with no evolution.  Combining the results, we
conclude that the ratio \mbh/\mstar\ rises with look-back time, although it
may saturate at $\approx 6$ times the local value.  Scenarios in which
moderately luminous quasar hosts at $z\gtrsim1.7$ were fully formed bulges
that passively faded to the present epoch are ruled out.
\end {abstract}


\section {INTRODUCTION}

A remarkable discovery over the last decade is the existence of strong
correlations between the masses \mbh\ of supermassive black holes (BHs) and
the overall properties of the host galaxy bulge, in particular the luminosity,
stellar mass (\mstar) and stellar velocity dispersion ($\sigma_*$) (e.g.,
Kormendy \& Richstone 1995; Magorrian et al.  1998; Ho 1999; Gebhardt et al.
2000a; Ferrarese \& Merritt 2000; Kormendy \& Gebhardt 2001, Marconi \& Hunt
2003, H\"aring \& Rix 2004).  There are also correlations with the narrow
[O~{\sc iii}] line width (Shields et al. 2003, Boroson 2003), and 
with the central stellar density profiles of the bulges (Graham et al.
2001).  The small intrinsic scatter, especially in the correlation between
\mbh\ and the host's stellar velocity dispersion, mass, and luminosity,
suggests that supermassive black hole growth and galaxy evolution are coupled
processes.  An empirical approach to understanding this coupling is to trace
these correlations to earlier epochs, determining whether the tight
correlations developed recently or long ago, and measuring the relative growth
rates of black holes and stellar bulges.  Unfortunately, it is impossible to
estimate BH masses beyond the local universe with the most robust tool --
spatially resolved kinematics.  Also, it is very difficult to measure stellar
velocity dispersions at redshifts $z\gtrsim 1$ in normal galaxies.

The most straightforward correlation to extend to high redshifts ($z\ge1$) is
the BH mass-bulge luminosity relation.  The correlation between \mbh\ and
galaxy luminosity, originally noted by Kormendy \& Richstone (1995) for
nearby, normal galaxies, has now been explored in great detail and measured in
passbands from the optical to the near infrared (e.g.  Magorrian et al.  1998;
Ho 1999; Kormendy \& Gebhardt 2001; Laor 2001; Merritt \& Ferrarese 2001;
Erwin, Graham, \& Caon 2002, McLure \& Dunlop 2002; Bettoni et al.  2003;
Marconi \& Hunt 2003, Ivanov \& Alonso-Herrero 2003).  These relationships are
fairly tight, with a typical scatter of 0.3-0.45 dex in BH mass for early-type
galaxies or early-type spiral galaxies that can be reliably decomposed into a
bulge and a disk (Erwin, Graham, \& Caon 2002, Bettoni et al. 2003; Marconi \&
Hunt 2003).  After two decades of ground-based and {\it HST} studies of quasar
hosts at low ($z\approx 0$) and intermediate ($z\le 1$) redshifts studies have
found that AGN hosts and normal galaxies share many of the same
characteristics (e.g.  Hutchings Crampton, \& Campbell 1984; Gehren et al.
1984; Neugebauer et al.  1985; Lehnert et al.  1992; McLeod \& Rieke 1994,
McLeod 1995; Lawrence 1999; Nelson 2000; Wang, Biermann, \& Wandel 2000):
they look alike in gross appearance.  The ubiquity of supermassive black holes
in galaxies suggests that most or all normal galaxies experience an AGN phase
at some point in their development.  And this AGN activity might itself have a
role in regulating star formation, which would affect the development of
early-type galaxies (Di Matteo, Springel, Herquist 2005; Ho 2005; Somerville
et al.  in prep.).

Studying the BH--bulge luminosity relation at high redshift requires measuring
both quantities in bright AGNs with comparatively faint hosts.  The AGN
activity allows \mbh\ to be estimated using the so-called virial technique
(see Ho 1999; Wandel, Peterson, \& Malkan 1999, Kaspi et al.  2000, McLure \&
Jarvis 2002, Vestergaard 2002, Peterson et al.  2004, Kaspi et al.  2005).
Reverberation mapping (Blandford \& McKee 1982; Peterson 1993) of nearby AGN
provides a correlation between broad-line region (BLR) sizes and AGN continuum
luminosity (Kaspi et al. 2000, Kaspi et al.  2005).  Combined with line
widths, this virial mass can is normalized to the local \msig\ relation to
provide estimates of \mbh.  These have been explored in detail locally
(Gebhardt et al.  2000b; Ferrarese et al.  2001; Nelson et al.  2004, Onken et
al. 2004; Barth, Greene, \& Ho 2005a; Greene \& Ho 2005, Kaspi et al.  2005)
and should work also at higher redshift.  What is harder to measure is the
luminosity of the host galaxy bulge; AGN activity and cosmological surface
brightness dimming make it difficult to detect distant AGN hosts, let alone
measure their properties.  To date, 70 orbits on {\it HST} have been used to
image just 15 objects at $z\approx 2$ in the NICMOS {\it H}-band (Kukula et
al.  2001, Ridgway et al.  2001).  Despite this modest sample size, it is
already possible to study the BH-bulge relationship, and Peng et al.  (2005,
P06 hereafter) show that these high redshift systems nearly lie on the local
\mbh\ versus rest-frame bulge luminosity relationship in the $R$-band.  If the
bulges at $z\simeq 2$ are fully formed and passively fading, this implies that
the mean \mbh/\mstar\ ratio is a factor of 3-6 higher at $z\gtrsim2$ than
today.  The amount of evolution is even larger if the hosts are analogous to
coeval inactive galaxies, which have lower mass-to-light (\ml) ratios than at
the present epoch due to their younger stellar populations (e.g.  Rudnick et
al.  2003).  This implies that initially black holes must have grown faster
than their bulges, with the bulges playing ``catch-up'' from $z\gtrsim2$ to
the present.  Indeed this evolution may still be observable in $z\approx 0.4$
Seyfert galaxies (Treu, Malkan, \& Blandford 2004), although the effect is
smaller.  The conclusions inferred from the high redshift study of P06,
however, were tentative because of the small sample size, covering two thin
redshift slices with only 15 objects.

Several other lines of evidence suggest that the host galaxies developed more
slowly at early epochs than their central black holes.  McLure et al.  (2005)
find similar evolution to the results of P06 by comparing {\it radio galaxy}
luminosities to {\it quasar} black hole masses.  There are risks, however, in
drawing inferences by comparing the properties of hosts and black holes in
potentially different types of objects.  Using CO lines to estimate the
stellar velocity dispersion $\sigma_*$ in high redshift quasar hosts, Shields,
Salviander, \& Bonning (2005) also find tentative evidence that the hosts are
undermassive compared to their central BHs.  The same conclusion is reached by
Merloni, Rudnick, \& Di Matteo (2004) who compared the star formation rate
density and stellar mass density in the universe with accretion rates of AGNs.
However, several other studies tell a conflicting story.  Using the [{\sc O
iii}] line width to approximate $\sigma_*$, Shields et al. (2003) find little
change in the \msig\ relation with redshift, albeit with a large scatter.
Finally, in contrast to all these studies, Borys et al.  (2005) find that the
\mbh/\mstar\ ratio may be $1-2$ order of magnitude {\it smaller} in high
redshift submillimeter selected galaxies than today, suggesting that BH masses
must grow rapidly from $z\approx 2$.  However, their finding is based on
assuming that their low luminosity AGNs are {\it all} radiating at the
Eddington limit, which places a lower limit on the BH mass.  In practice, the
Eddington ratios in AGNs are observed to span anywhere from near 0 for normal
galaxies to unity (Woo \& Urry 2002), and there are abundant examples of gas
rich galaxies that have no detectable AGN.  Because of the complications of
these other techniques, the most promising way to quantify the evolution of
\mbh/\mstar\ is still to measure \mbh\ and the host luminosity in the same
objects.  To do so, however, requires a much larger sample than is currently
available in the literature.

\begin{figure} 
    \centerline {\includegraphics[angle=0,width=9cm]{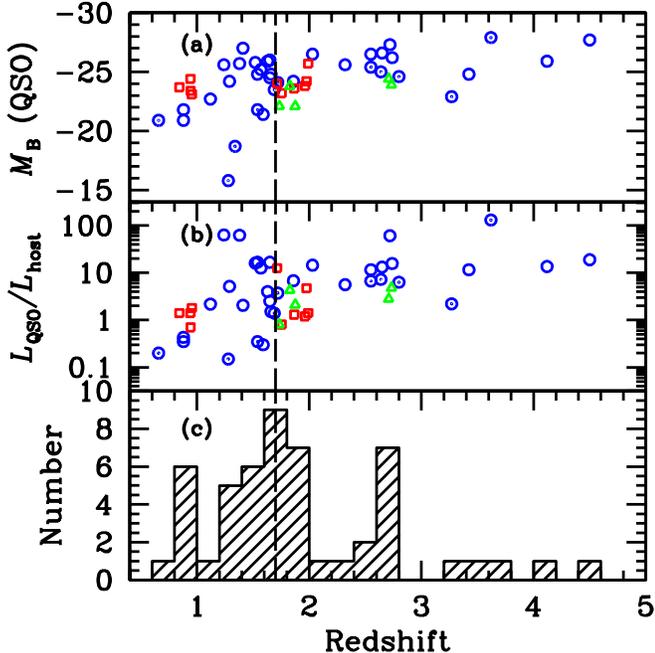}}
    \caption {Summary of the quasars used in this sample.  We include
	      gravitational lenses from the CASTLES project (circles are
	      2-image lenses; circle-dot are 4-image lenses), and from direct
	      NIC2 imaging studies by Kukula et al.  (2001, red squares) and
	      Ridgway et al. (2001, triangles).  The dashed line at $z=1.7$
	      separates our high and low redshift samples into two bins, each
	      covering approximately 2~Gyrs of look-back time.  Panel {\it a}
	      shows the $B$-band absolute magnitude of the quasar, {\it b}
	      shows the $H$-band ratio between the quasar and host
	      luminosities, and {\it c} shows the redshift histogram.}
    \label{fig:zdist}
    \vskip 0.15truein
\end {figure}

Here we improve on the number statistics by using gravitationally lensed host
galaxies to expand our analysis to 51 objects (see Fig.~\ref{fig:zdist}).
Gravitational lensing stretches the host galaxy out from underneath the quasar
emission, preserving the surface brightness of the host but stretching and
distorting its shape, leading to an enormous gain in surface brightness
contrast between the host and the quasar compared to unlensed systems.  The
distortion of the host into an Einstein ring also gives the image of the host
a dramatically different morphology from that of any errors in the point
spread function (PSF), so the photometry of the host is significantly less
susceptible to systematic problems in the PSF model.  Thus, while Kukula et
al.  (2001) and Ridgway et al.  (2001) found it challenging to quantify the
structure of $z\simeq 2$ hosts even with five-orbit {\it HST} images, many
single orbit images of lensed hosts are easily analyzed because the hosts are
observed as luminous arcs and Einstein rings on scales of 1--10~arcsec (see
Figures~\ref{fig:he1104}--\ref{fig:bri0952}; also Keeton et al.  2000,
Kochanek et al. 1999, Leh\'ar et al. 2000).  Given the existence of $\sim 80$
lensed quasars, {\it HST} images of lensed quasars are by far the most
efficient way of expanding studies of quasar host galaxies, especially at
$z\approx 2-4$.


In the present study, we measure the rest-frame $R$-band (\lr) luminosities of
31 lensed + 5 unlensed host galaxies from the CASTLES (CfA-Arizona Space
Telescope Lens Survey) project, and combine the results with the sample of 15
quasars from Kukula et al. (2001) and Ridgway et al.  (2001) that we discussed
in P06.  We estimate \mbh\ for the sources using the virial method, and then
explore the evolution of the ratio between \mbh\ and both the rest frame
$R$-band bulge luminosity \lr\ and the estimated stellar mass
\mstar\ between $z\simeq 1$ and $z\simeq 4.5$ given conservative models for
the stellar populations and their evolution.  In Section~\ref{sect:data} we
describe the lens data and analysis, summarize the P06 sample, and present our
approach to making {\it k}-corrections.  In Section~\ref{sect:virial} we
estimate the BH masses using the virial technique.  Section~\ref{sect:results}
presents the results.  Finally, we conclude in Section~\ref{sect:conclusion}
with a discussion of the evolution of the \mbh/\mstar\ ratio from $z\simeq
4.5$ to today.  An Appendix discusses systematic uncertainties and explains
why they do not affect our conclusions.  In particular, we discuss assumptions
made about the host galaxy stellar population and evolution (Appendix
\ref{app:assumptions}), the accuracy of the lens models
(\ref{app:degeneracies}), their complexity (\ref{app:complexity}), and lastly,
specifics on several problematic objects (\ref{app:problematic}).  All the
results assume a standard cosmology with $H_0=70$ km$^{-1}$ s$^{-1}$
Mpc$^{-1}$, $\Omega_{\rm m} = 0.3$, and $\Omega_\Lambda = 0.7$.  The
photometry is in the Vega system and is corrected for Galactic extinction
using the models of Schlegel, Finkbeiner \& Davis (1998).

\section {GRAVITATIONAL LENS DATA AND MODELING}
\label {sect:data}

In this paper we analyze the host galaxies of lensed quasars observed with
{\it HST} as part of the CfA/Arizona Space Telescope Lens Survey (CASTLES),
focusing on those that have both NICMOS/NIC2 $H$-band (F160W)
images\footnote{The $H$-band zeropoints are $H_{\rm zpt}=21.79$ for data
obtained prior to year 2000, and $H_{\rm zpt}=22.11$ post 2000.}, and
published spectra for estimating BH masses.  The $V$-band (F555W) and $I$-band
(F814W) images obtained as part of CASTLES reflect the rest-frame near- to
far-UV radiation and will only be used in a later study.  We restricted the
analysis to lenses with Einstein ring diameters exceeding $0\farcs8$ in order
to simplify modeling the images to determine the properties of the host
galaxy.  This choice corresponds to setting a minimum mass for the lens
galaxy, and should introduce no biases in the properties of the host galaxy of
the quasar source.  The data were reduced as detailed by Leh\'ar et al.
(2000).  For most targets we have only single orbit observations, but a few
sources have longer, five orbit observations.

Table 1 summarizes the CASTLES dataset.  While this sample includes 31 lensed
objects there are 5 which are unlensed: the host galaxy of quasar C in the
field of the lensed quasar LBQS~1009--0252 (Hewett et al. 1994); two binary
quasar pairs MGC~2214+3550 (Mu\~noz et al. 1998) and RXJ~0921+4258 (Mu\~noz et
al.  2001).  We have reclassified this last system as a binary quasar based on
the properties of the host galaxy (see Appendix~\ref{app:problematic}).  We
complemented this sample by including the unlensed 15 objects from Kukula et
al.  (2001) and Ridgway et al.  (2001), summarized in Table 2.  Their
estimated $H$-band fluxes are published previously (Kukula et al. 2001,
Ridgway 2001) and corrected by P06 for morphological corrections and
extinction.  The full sample consists of 51 hosts spread over the redshift
range 1--4.5.  We will present a survey of the host galaxy morphologies and
colors in a separate study.

We deliberately excluded observations of unlensed host galaxies for which the
only available data were at optical rather than near-IR wavelengths (e.g.,
Aretxaga, Terlevich, \& Boyle 1998; Lehnert et al. 1999; Hutchings et al.
2002; Hutchings 2003, S\'anchez et al. 2004, Jahnke et al 2004).  These
(rest-frame) ultraviolet observations are very sensitive to star formation
rates and extinction internal to the host galaxy, which makes $k$-corrections
very uncertain, and they are not easily compared to local correlations.  We
also excluded ground-based near-infrared results (e.g. Falomo, Kotilainen \&
Treves 2001, Lacy et al.  2002, S\'anchez \& Gonz\'alez-Serrano 2003, and
Kuhlbrodt et al. 2005) in favor of a uniform analysis based on {\it HST}
observations.

\subsection{Image Modeling}


\begin{figure*} 
    \centerline {\includegraphics[angle=0,width=18.0cm]{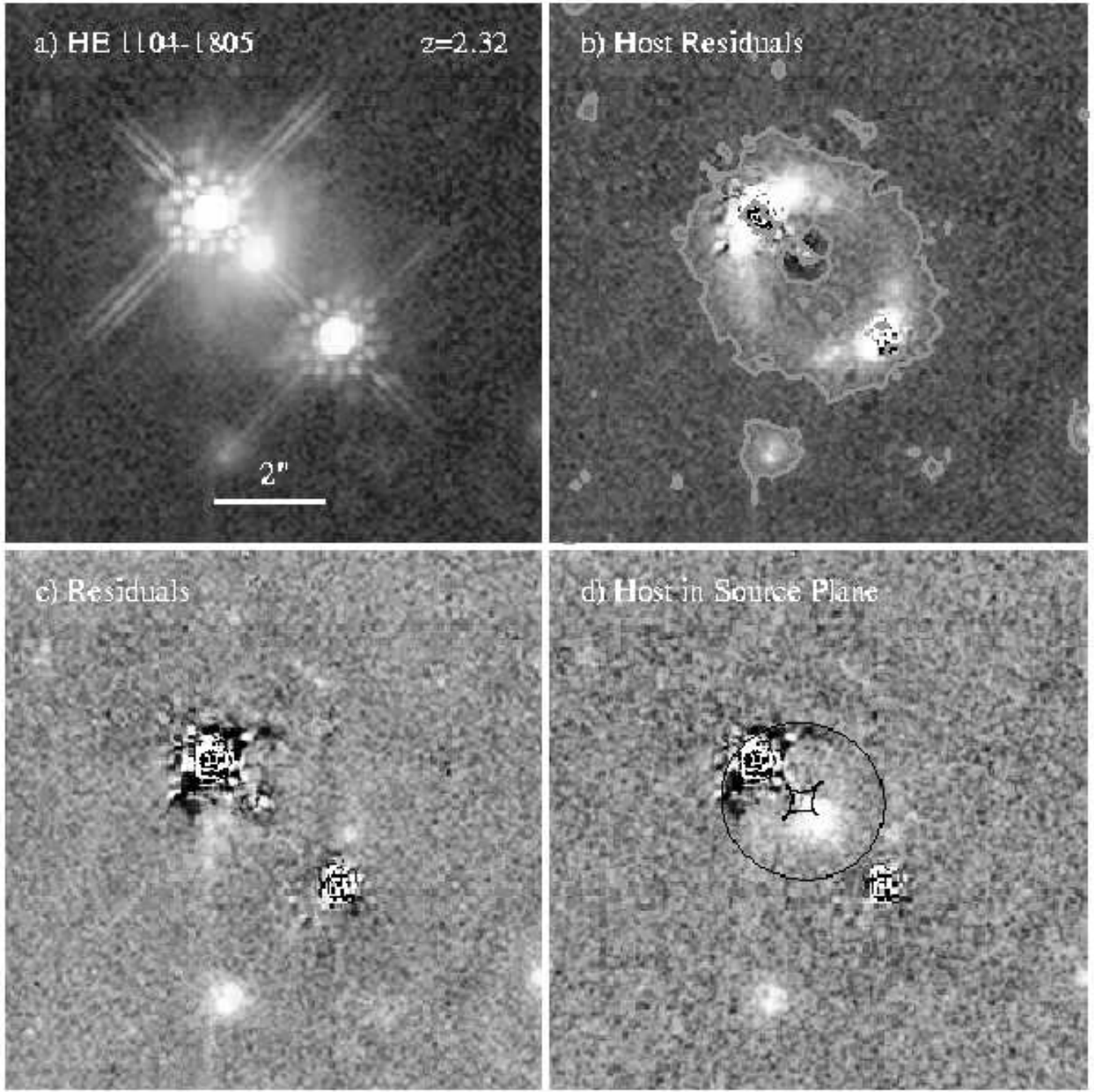}}
    \caption {The two image lens HE~1104$-$1805 of a $z=2.32$ quasar produced
	      by a $z=0.73$ lens galaxy.  Panel~{\it a} shows the original
	      data, Panel~{\it b} shows the lensed host galaxy found after
	      subtracting the lens and quasar components of the best fitting
	      photometric model, Panel~{\it c} shows the residuals from that
	      photometric model, and Panel~{\it d} shows what the unlensed
	      host galaxy would look like in a similar exposure after
	      perfectly subtracting the flux from the quasar.  In this 5 orbit
	      exposure, the host, which we estimate to be a galaxy with
	      concentration $n=2$ and effective radius $r_e=2.5$~kpc, is
	      observed as a complete Einstein ring.  The curves shown
	      superposed on the model of the host galaxy are the lensing
	      caustics.}
    \label{fig:he1104}
    \vskip 0.15truein
\end{figure*}

\begin{figure*} 
    \centerline {\includegraphics[angle=0,width=18.0cm]{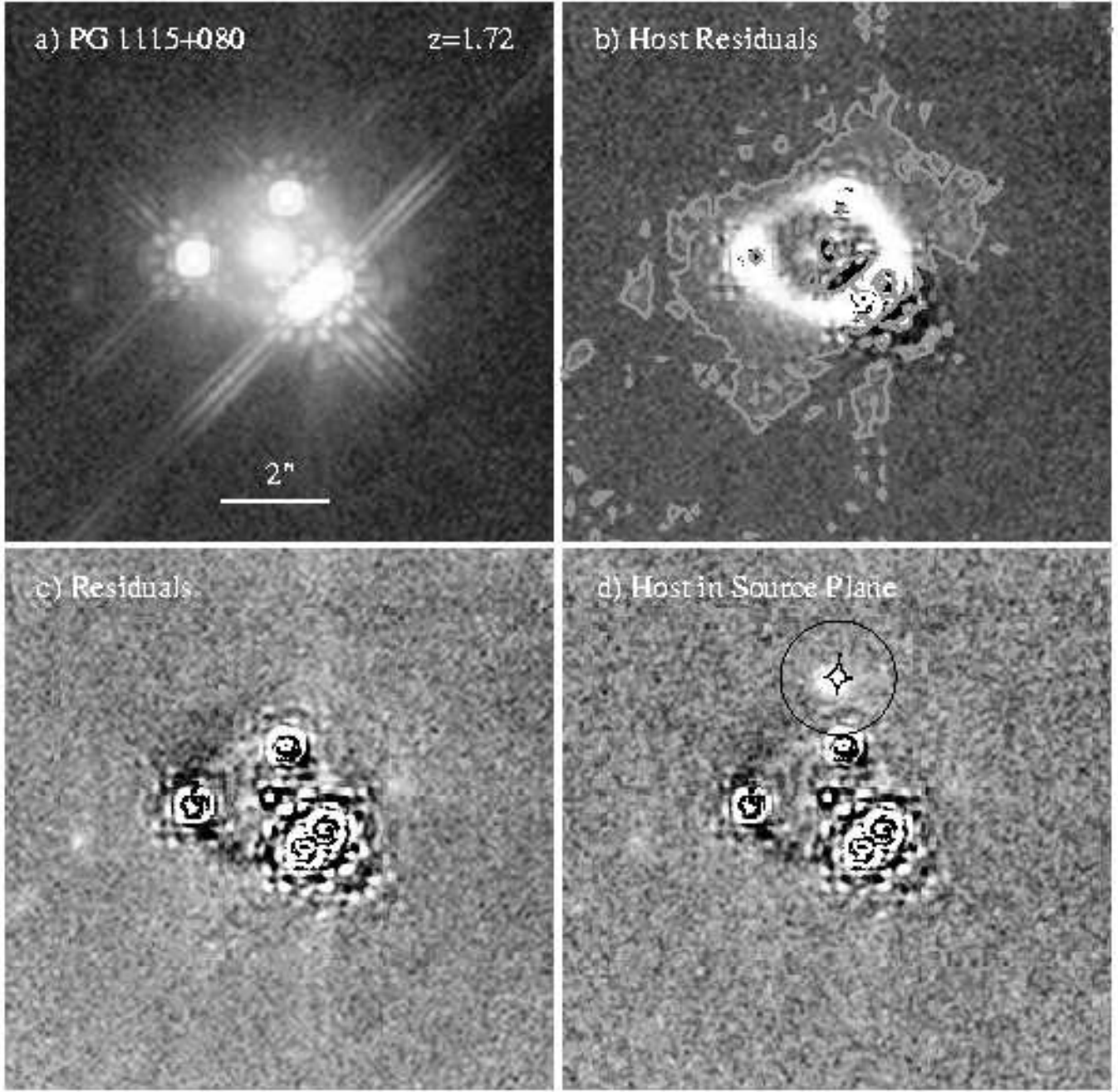}}
    \caption {The four image lens PG~1115+080 of a $z=1.72$ quasar produced
	      by a $z=0.31$ lens galaxy.  The panels are described in
	      Fig.~\ref{fig:he1104}.  In this case we estimate that the host
	      has a concentration of $n=4$ and an effective radius of $r_e
	      \simeq 1.5$~kpc.  The large apparent spatial displacement
	      between the caustics from the center of the lensing galaxy is
	      due the gravitational influence of a neighboring galaxy external
	      to the field of view, whose lensing effect is represented by a
	      singular isothermal sphere mass model (Impey et al. 1998).}
    \label{fig:pg1115}
    \vskip 0.15truein
\end{figure*}

\begin{figure*}
    \centerline {\includegraphics[angle=0,width=18.0cm]{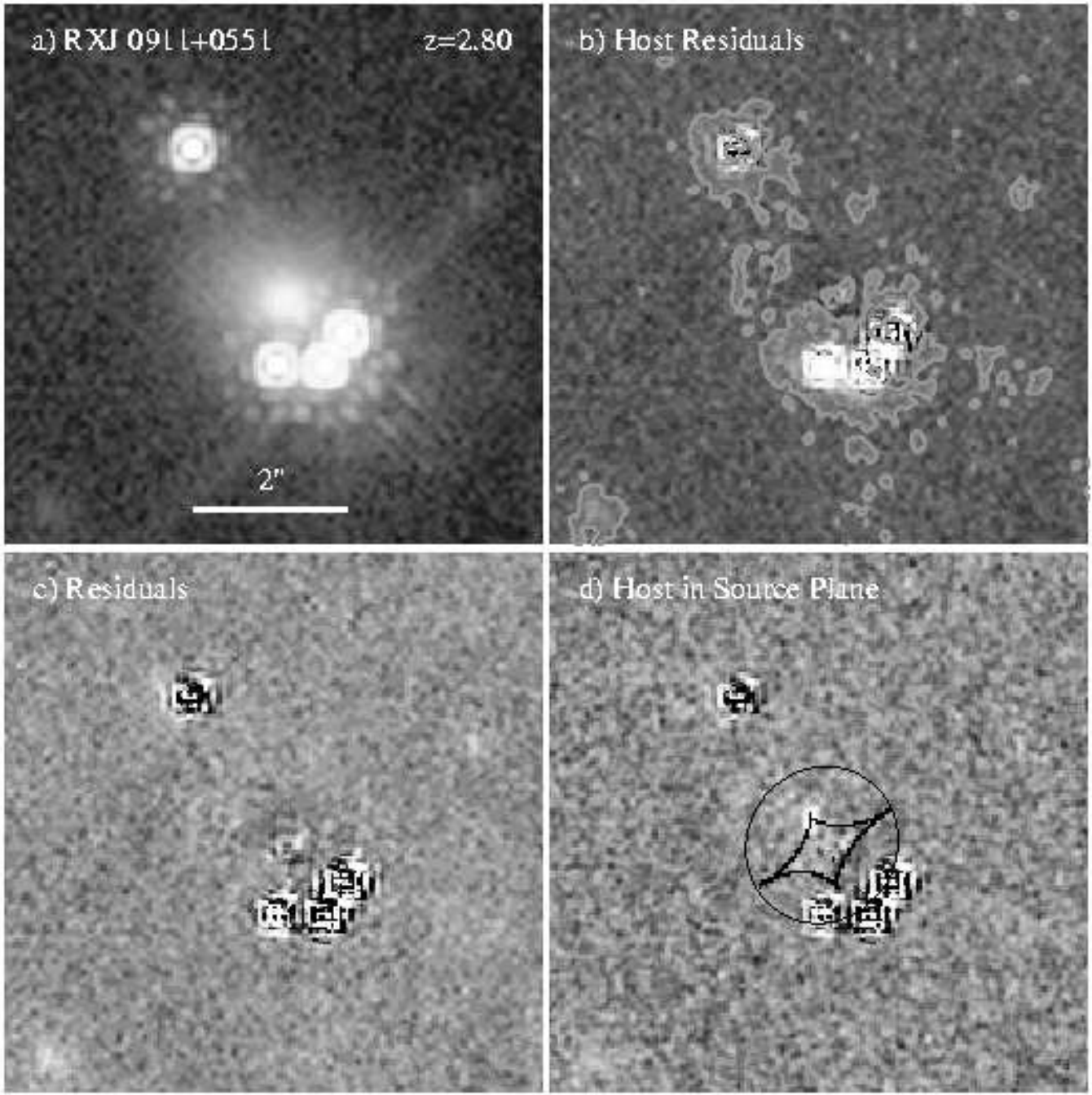}}
    \caption {The four image lens RXJ~0911+0551 of a $z=2.8$ quasar
	      produced by a $z=0.77$ lens galaxy.  The panels are described in
	      Fig.~\ref{fig:he1104}.  In this case we estimate that the host
	      has an effective radius of $r_e \simeq 1-2$~kpc..}
    \label{fig:rxj0911}
    \vskip 0.15truein
\end{figure*}

\begin{figure*}
    \centerline {\includegraphics[angle=0,width=18.0cm]{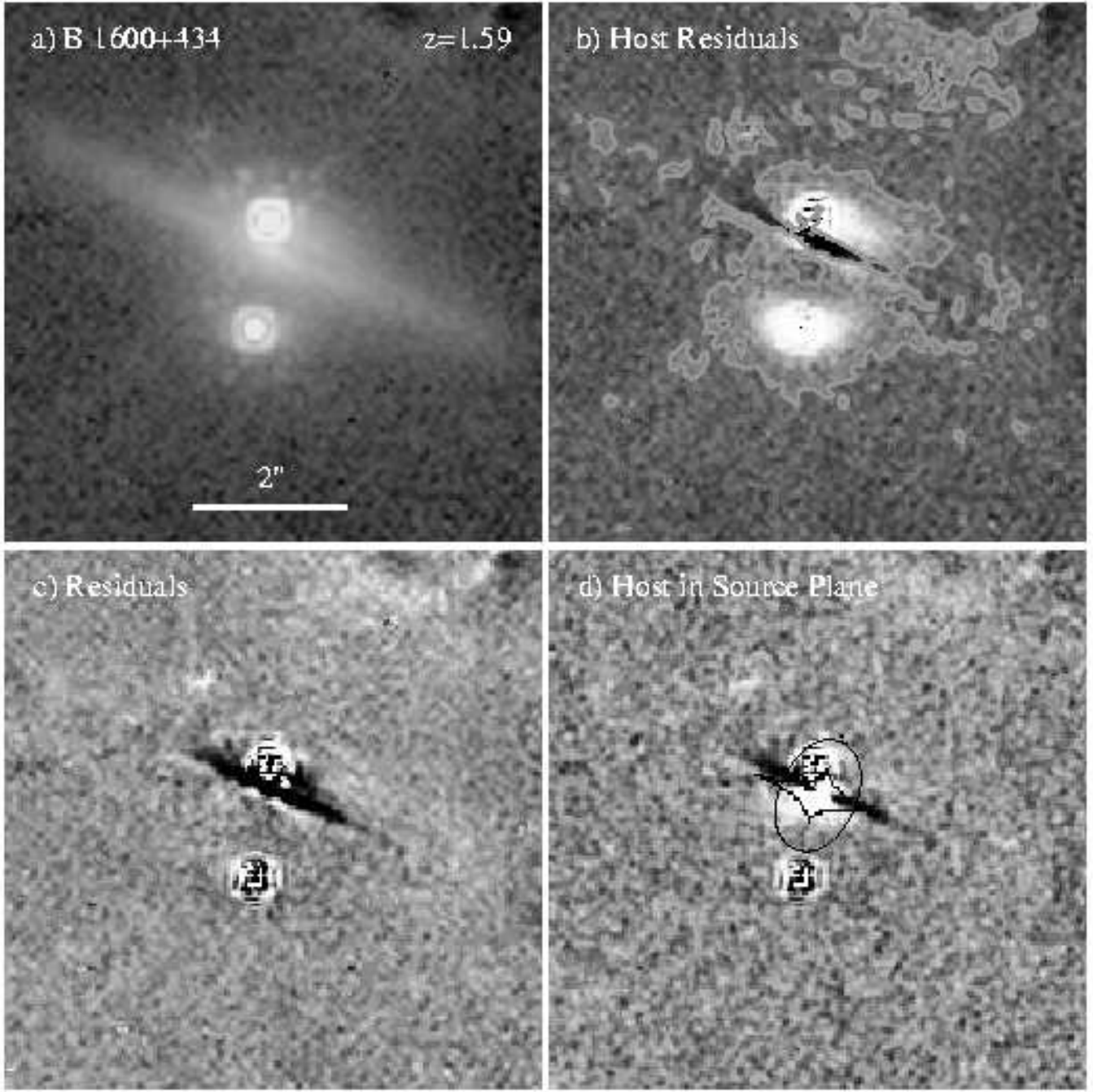}}
    \caption {The two image lens B~1600+434 of a $z=1.59$ quasar produced by
	      a $z=0.41$, edge-on spiral lens galaxy.  The panels are
	      described in Fig.~\ref{fig:he1104}.  In this case we estimate
	      that the host has an effective radius of $r_e \simeq 3$~kpc.}
    \label{fig:b1600}
    \vskip 0.15truein
\end{figure*}

\begin{figure*}
    \centerline {\includegraphics[angle=0,width=18.0cm]{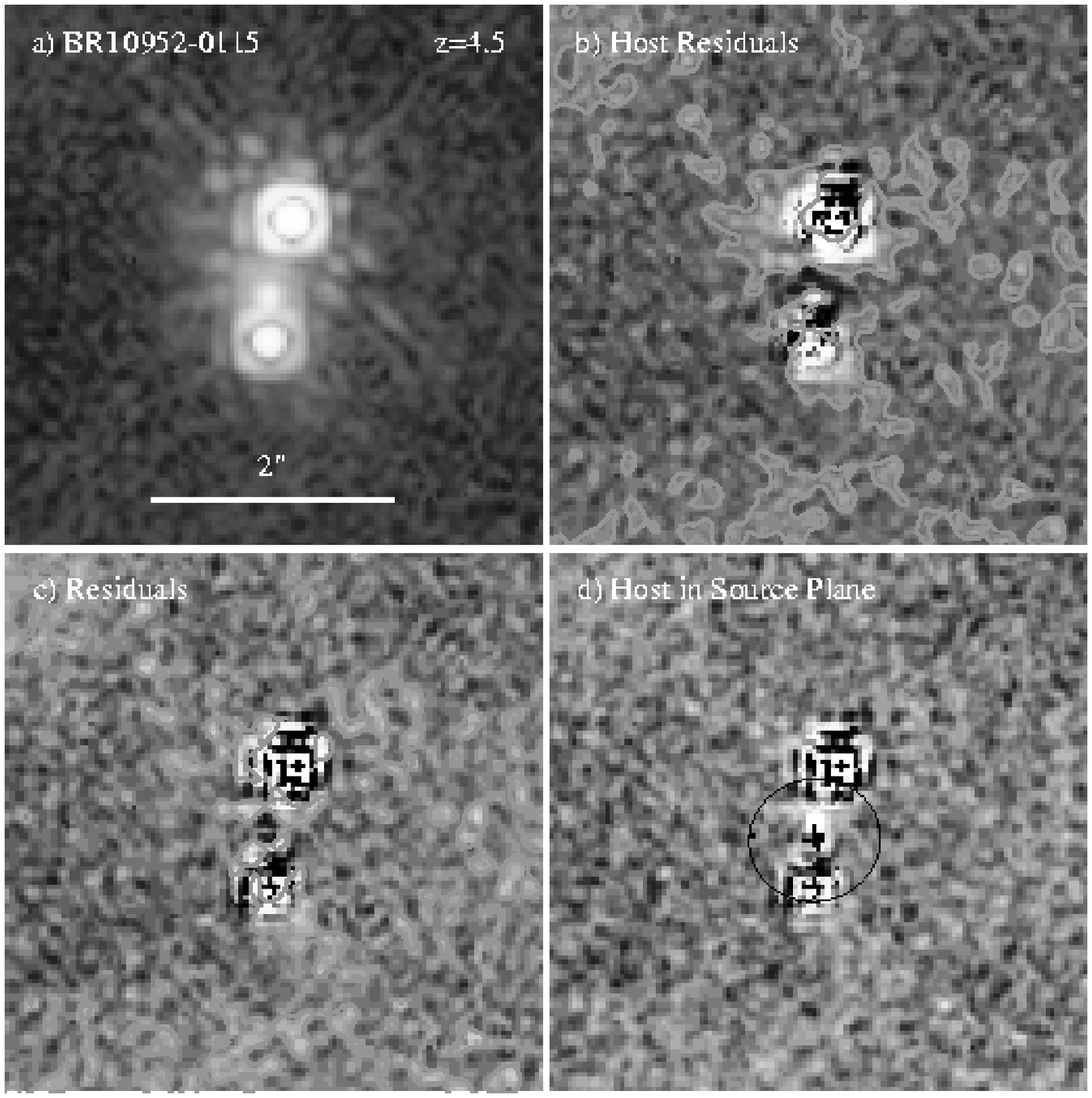}}
    \caption {The two image lens BRI~0952$-$0115 of a $z=4.5$ quasar
	      produced by a $z\approx0.41$ lens galaxy.  The panels are
	      described in Fig.~\ref{fig:he1104}.  The host is resolved into
	      an arc.  In this case we estimate that the host has an effective
	      radius of $r_e \lesssim 1$~kpc.}
    \label{fig:bri0952}
    \vskip 0.15truein
\end{figure*}


We determine the host galaxy properties given in Tables~1 and 2 by fitting a
model to the image using either GALFIT (for unlensed sources, Peng et al.
2002) or LENSFIT (for lensed sources).  LENSFIT is a version of GALFIT that
has been extended to fit lensed host galaxies while optimizing the mass model
for the lens galaxy.  Briefly, a LENSFIT model comprises S\'ersic (1968)
models for the lens galaxy and quasar host galaxy, and point images of the
quasar.  The quasar and host are distorted by a lens mass model consisting of
one or more singular isothermal ellipsoids (SIE) in an external shear field.
The model is convolved with the point spread function (PSF) and fitted to the
observations.  All the model parameters, both for the luminous components and
the lens mass model, are simultaneously optimized using the
Levenberg-Marquardt non-linear least squares method as implemented by Press et
al. (1992).  The parameter uncertainties include all covariances among model
parameters.  As discussed in the Appendix, lensing introduces no significant
systematic uncertainties in the measurements of the quasar host galaxies.

We model the light profiles of both the foreground lensing galaxies and
the background host galaxies (the ``source'') using the S\'ersic (1968) model,
\begin {equation}
  \Sigma(r)=\Sigma_e 
  \exp\left(-\kappa(n) \left[\left({\frac{r}{r_e}}\right)^{1/n} - 1\right]\right).
\end {equation}
\noindent The S\'ersic profile is widely used because it reliably models
galaxy light profiles and smoothly covers the range of profile shapes from
exponential disks ($n=1$) to de~Vaucouleurs profiles ($n=4$). The scale length
$r_e$ is defined to be the half-light radius of the profile, so the
coefficient $\kappa(n)$ is the function of concentration $n$ needed to
maintain this definition (see Peng et al. 2002 for details).  We use
elliptical S\'ersic profiles, so there are a total of 7 parameters: the
position, the flux scale $\Sigma_e$, the half-light radius $r_e$, the
concentration $n$, the axis ratio $q$ and the position angle $\theta$ of the
major axis.  In the absence of lensing, this procedure is identical to the
approach used in many studies of host galaxies (e.g.  McLure et al.  1999,
McLeod \& McLeod 2001, Peng et al.  2002, Dunlop et al. 2003, S\'anchez et al.
2005, Jahnke et al.  2005).

Like all other studies of quasar hosts, fitting the data requires a good model
for the PSF.  However, the details are much less important to lensed hosts
than unlensed hosts, because lensed hosts are much more extended and have
structures that differ dramatically from the PSF.  For example, small errors
in the width of the PSF are an enormous problem for analyzing unlensed hosts
but have almost no effect on the estimated properties of a host seen as an
Einstein ring.  This is an important advantage of using lensed hosts -- even
though {\it HST} produces very stable PSFs, matching them to quasars is
difficult because quasars and stars have different spectral shapes in the
infrared.  We test for PSF-related problems and biases for the parameters of
hosts near our detection thresholds by doing the fits using a collection of 13
high signal-to-noise images of stars.  We fit each system with all PSF models,
presenting the results for the PSF producing the best fit to the data.

The one challenge of this study, and the main difference from unlensed
studies, is that the apparent geometry of the host galaxy and quasar images
must be reproduced by a gravitational lens model.  As a problem in image
fitting, this is merely an inconvenience, since all the lens does is to remap
the ``unlensed'' source plane coordinates $\vec{u}$ onto the ``lensed'' lens
plane coordinate $\vec{x}$.  On the source plane the host galaxy can again be
modeled by an analytic profile $\Sigma(\vec{u})$.  Since gravitational lensing
conserves surface brightness, the surface brightness of the lensed host is
simply $\Sigma(\vec{x})=\Sigma(\vec{u})$.  Given a model for the lens
potential $\phi(\vec{x})$ in appropriate units, the lens equation,

\begin{equation}
\label {lenseq}
     \vec{u} = \vec{x} - \vec{\nabla}_{\vec{x}} \phi(\vec{x}).
\end{equation}

\noindent gives the mapping from source to lens plane coordinates (Schneider,
Ehlers, \& Falco 1992).  The remapping requires a lens model.  We outline our
modeling procedure here, leaving more detailed discussion to a later study
(Peng et al.  in preparation).  We model the lens as one or more singular
isothermal ellipsoids (SIE) in an external (tidal) shear field.  The SIE has a
projected mass density of

\begin {equation}
\kappa(x,y) = {\frac{b}{2}} \left[{\frac{2q^2}{1+q^2}} \left(x^2+y^2/q^2\right)\right]^{-1/2},
\end {equation}

\noindent specified by a mass scale $b$, an axis ratio $q$ and a major axis
position angle.  The parameter $b$ is approximately the radius of the Einstein
ring the lens will form, although the relation is exact only for $q=1$ (see
Kochanek, Keeton, \& McLeod 2001).  The tidal shear, as its name suggests, is
needed to model the tidal effects from objects near the lens or along the line
of sight to the quasar (Keeton, Kochanek, \& Seljak 1997).  The parameters of
the lens model are adjusted as part of fitting the data.

For the types of systems we consider, the geometry of the lensed images and
the host galaxy constrain the parameters of the lens model quite accurately.
Moreover their uncertainties have little effect on the conclusions, and the
lensed images are well-fit by the model.  Indeed, the random and systematic
uncertainties of the host photometry due to PSF issues are significantly
larger than uncertainties due to the lens model.  This, combined with the
diminished importance of the character of the PSF, is what give lensing its
much greater sensitivity to hosts relative to direct imaging.  To reassure
readers unfamiliar with lens models, we discuss this further in
Appendices~\ref{app:degeneracies} and \ref{app:complexity}.

Figure~\ref{fig:zdist} summarizes the luminosity, quasar/host contrast and
redshift distribution of the full sample.  In general, the unlensed quasars of
Kukula et al.  (2001) and Ridgway et al.  (2001) were selected to be lower
luminosity AGNs, while the lenses are a heterogeneous sample drawn from a
broad range of optical- and radio-selected lens surveys with differing
sensitivities.  The benefit of the lensing magnification is easily apparent in
Figure~\ref{fig:zdist} --- we can probe to higher average quasar luminosities
and quasar/host galaxy contrast levels than with the unlensed quasars.  We can
also measure the concentration index $n$ of roughly 80\% of the lensed hosts
(the other 20\% are held fixed to $n=4$), despite the relatively shallow data
(1-orbit) available for most of the objects (Peng et al. 2006, in
preparation).  We have size estimates for all lensed hosts with luminosity
measurements, whereas they are poorly constrained in the absence of lensing,
despite their longer integration times (Kukula et al. 2001, Ridgway et al.
2001).

Figures~\ref{fig:he1104}--\ref{fig:bri0952} illustrate five examples of
lensed hosts: HE~1104$-$1805 ($z_s=2.32$), PG~1115+080 ($z_s=1.72$),
RXJ~0911+0551 ($z_s=2.80$), B~1600+434 ($z_s=1.59$), BRI~0952$-$0115
($z_s=4.5$).  With the exception of the one-orbit observation of
BRI~0952$-$0115, these are all examples of the deeper five-orbit observations.
For each lens we show the raw data, the structure of the lensed host after
using the best fit model to subtract the contributions from the lens galaxy
and the quasar point sources, and the final residuals from subtracting the
full model.  We also show a model, including noise, of what the host would
look like in a comparable HST image if it had not been gravitationally lensed
and assuming perfect subtraction of the quasar point source flux.  The lensed
hosts not only gain from a net magnification of their flux (by factors of 8,
18, 7, 4 and 8 for HE~1104$-$1805, PG~1115+080, RXJ~0911+0551, B~1600+434, and
BRI~0952$-$0115, respectively) but from an even more dramatic {\it contrast
enhancement} due to being stretched out from underneath the unresolved quasar
by the magnification.  We discuss some of these issues further in the
Appendices.

\subsection {Estimating The Rest-Frame $R$-Band Luminosities of Host Galaxies}

\label{subsect:kcorr}

\begin{figure}
    \centerline {\includegraphics[angle=0,width=9.0cm]{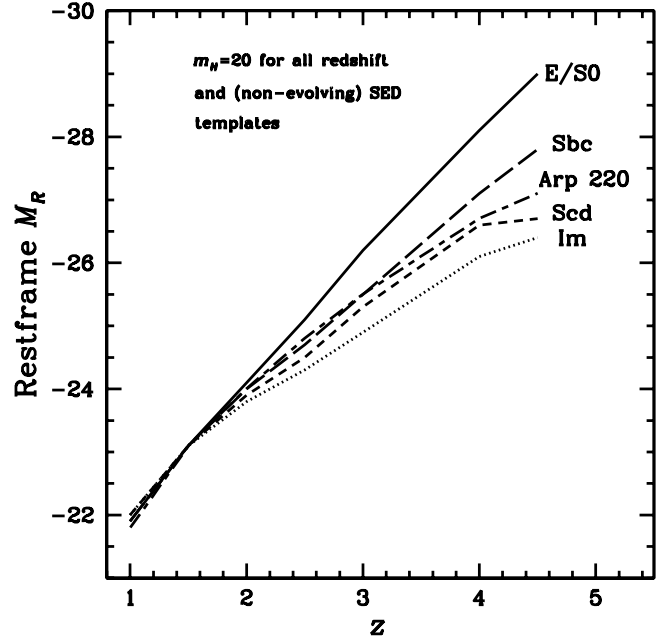}}
    \vskip -0.1truein
    \caption {The absolute, restframe $R$-band luminosity for a galaxy with
	      an observed $H$-band luminosity of 20~mag for four different
	      models of the SED.  The SEDs are the E/S0, Sbc, Scd, and Im
	      models from Coleman, Wu, \& Weedman (1980).  Note that for our
	      redshift range, all SEDs bluer than the E/S0 model have fainter
	      restframe luminosities than the estimate from the E/S0 model.}
     \label{fig:sedmags}
     \vskip 0.15truein
\end {figure}

To proceed further, we use the $H$-band flux measurements to estimate the
rest-frame $R$-band luminosities of the host galaxy bulges presented in Tables
3 and 4.  Fig.~\ref{fig:sedmags} shows the conversion between $H$-band and
rest-frame $R$-band as a function of redshift for galaxies with different
spectral types from Coleman, Wu \& Weedman (1980).\footnote{We follow Hogg et
al.  (2004) in computing detailed $k$-corrections.  We transform the {\it HST}
Vega-based magnitudes to $R$-band using the Coleman, Wu, \& Weedman (1980)
templates and appropriate filter transmission curves.  The integrated fluxes
are normalized to a spectrum of $\alpha$-Lyr, ``observed'' in the appropriate
bandpass.} Since the $H$-band roughly corresponds to the rest frame $R$-band
at $z\sim 1.4$ the conversions are only weakly dependent on the assumed SED
for the redshift range $1 \lesssim z \lesssim 2.5$.  For example, at $z=2$ the
spread between doing the conversion with the reddest (E/S0) and bluest (Im)
SEDs is only 0.3~mag (Figure~\ref{fig:sedmags}).  At higher redshifts, the
choice of SED becomes very important, and by $z\sim4.5$ the magnitude
difference varies by 2.5~mag between the E/S0 and Im SED models.  We adopt the
Sbc template as our standard model.  The Sbc template predicts an $I-K$ color
of $\simeq 3$~AB mag that is typical of a distant, $z\gtrsim 2$ red galaxy
(Labb\'e et al. 2005).  It is 1--2 mag redder than typical Lyman break
galaxies, which are better modeled by the Im SED.  While we adopt the Sbc
model as our standard model, we will explore the consequences of the extreme
assumptions of an E/S0 or Im SED on our results.

One extreme alternative to using a $z=0$ Sbc SED for the $k$-correction
template is to use that of an Ultra-Luminous Infrared Galaxy (ULIRG) because
the high levels of star formation and dust obscuration in a ULIRG may be more
representative of quasar host galaxies than a nearby normal galaxy.  As shown
in Fig.~\ref{fig:sedmags} the $k$-corrections derived from the Devriendt,
Guiderdoni, \& Sadat (1999) SED for the nearby, prototypical ULIRG Arp 220 are
midway between those of the Sbc and Scd templates at all relevant redshifts.
Thus, our estimated restframe luminosities are insensitive to whether we treat
the host as a normal galaxy or a ULIRG.

Since the lensed quasar and host images pass near or through the lensing
galaxy, some extinction may be caused by the lensing galaxy.  Typical lens
galaxies produce differential extinction of lensed quasars with a median
$\Delta E(B-V)\simeq 0.08$~mag (Falco et al.  1999).  For a standard $R_V=3.1$
extinction law this would lead to a correction to the host luminosity
(assuming a uniform dust screen) of only $A_H \simeq 0.05$~mag for a lens at
$z=1$, with smaller corrections at lower lens redshifts.  There is
considerable evidence that the lens galaxies show a range of extinction curves
(e.g. Falco et al. 1999, Mu\~noz et al.  2004), but these variations occur at
wavelengths $\lesssim 4000$~\AA\ rather than the rest-frame near-IR or optical
wavelengths that we use.  There are some lenses with significantly higher
extinction (in B1608+656 differential extinction of the host galaxy is
observed -- Koopmans et al.  2003; Surpi \& Blandford 2003), but we have not
included any of these systems in our analysis.  In short, we make no
corrections for extinction of the host galaxy by dust in the lens galaxy
because they are likely to be small.

\section {VIRIAL BLACK HOLE MASS ESTIMATE}
\label {sect:virial}

We estimate \mbh\ using the virial technique applied to the C~{\sc iv}
($\lambda$~1549~\AA), Mg~{\sc ii} ($\lambda$~2798~\AA), and the H$\beta$
($\lambda$~4861~\AA) emission line widths and their local continuum
luminosities $\lambda L_\lambda(1300~{\mbox{\AA}})$, $\lambda
L_\lambda(3000~{\mbox{\AA}})$ and $\lambda L_\lambda(5100~{\mbox{\AA}})$
respectively.  We adopt the normalization of Onken et al. (2004), confirmed
through a much larger sample by Greene \& Ho (2005), between the virial
relations to local measurements of black hole masses.  This normalization is a
factor of 1.8 higher than earlier normalizations.  Combining the BLR radius
estimates of Kaspi et al. (2005) and the new normalization factor, we adopt
virial relations of

\begin {equation}
{\mathcal M}_{\rm BH} = A_{\rm{line}}\left[\frac{{\rm FWHM(\rm{line})}}{1000\ {\rm km\ s}^{-1}}\right]^2\left[\frac{\lambda L_\lambda(\lambda_{\rm{line}})}{10^{44}\ {\rm erg\ s}^{-1}}\right]^{\gamma_{\rm{line}}} \, {\mathcal M}_{\odot},
\end {equation}

\noindent with $\lambda_{\rm{line}}\{\mbox {\sc Civ},\mbox{Mg~\sc
ii},\mbox{H~$\beta$}\} = \{1350, 3000, 5100\}$~\AA, $A_{\rm{line}}\{\mbox
{\sc Civ},\mbox{Mg~\sc ii},\mbox{H~$\beta$}\} = \{4.5, 6.1, 5.9\}\times10^6$,
and $\gamma_{\rm{line}}\{\mbox {\sc Civ},\mbox{Mg~\sc ii},\mbox{H~$\beta$}\}
= \{0.53, 0.47, 0.69\}$ (Vestergaard \& Peterson 2006, McLure \& Jarvis 2002, Kaspi
et al. 2005, respectively).  We refer the reader to P06 and references therein
for a detailed discussion of the method and its limitations.

\begin{figure*}
    \centerline {\includegraphics[angle=-90,width=18.0cm]{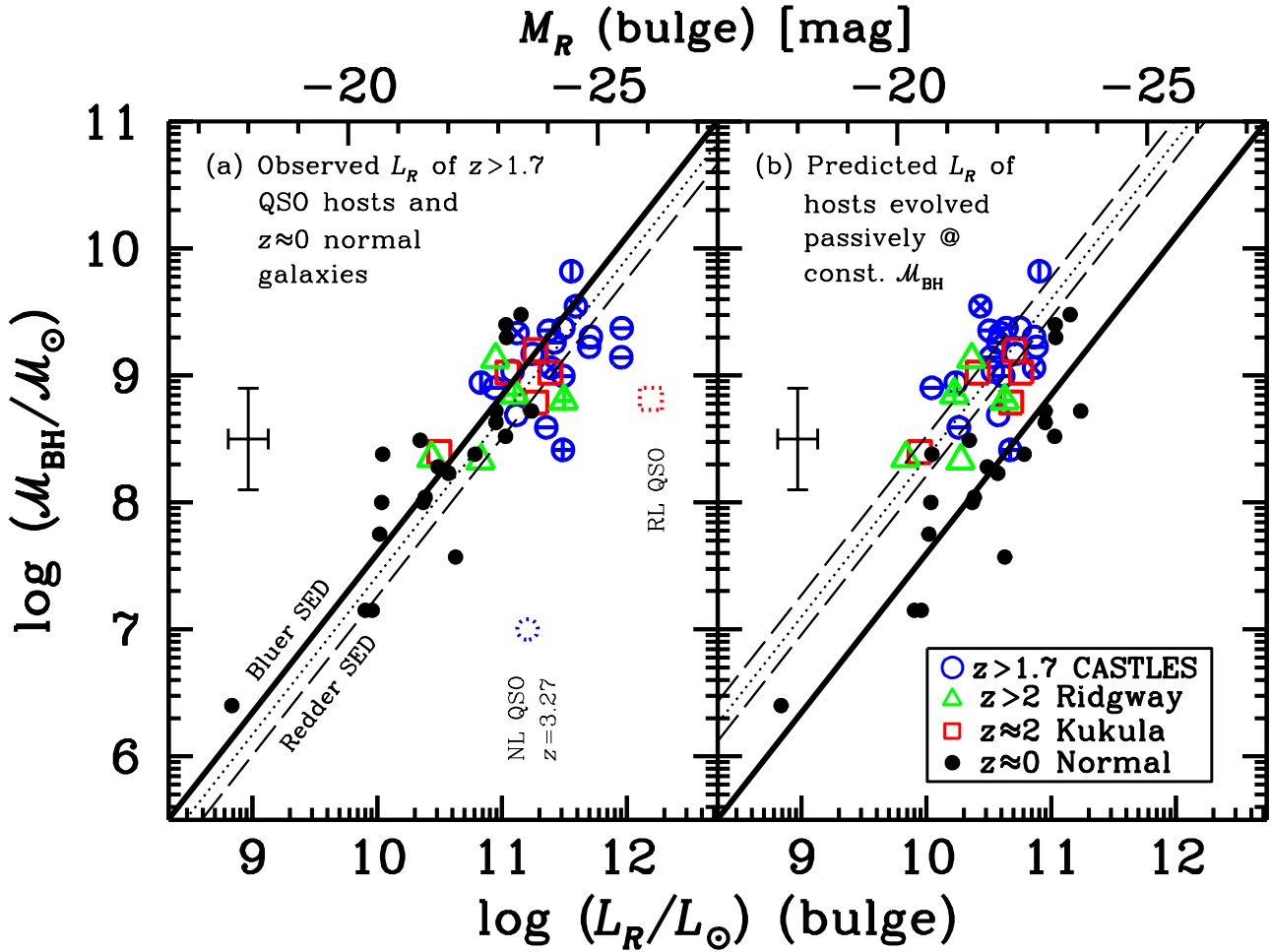}}
    \vskip -0.1truein
    \caption {The observed ({\it a}, left) and evolution corrected ({\it
	      b}, right) correlations between black hole mass and host
	      luminosity for the high redshift host sample.  The round solid
	      points are the local comparison sample, and the solid line is
	      the best fit, local relationship between \mbh\ and host
	      luminosity.  The high redshift hosts are shown using circles for
	      the gravitationally lensed hosts, triangles for the Ridgway et
	      al.  (2001) hosts and squares for the Kukula et al. (2001)
	      hosts.  An Sbc SED was used to make all the $k$-corrections.
	      The dotted line shows the best fit relation for the high
	      redshift hosts using the Sbc SED to make the $k$-corrections,
	      and the dashed lines show the effect of using either the E/S0
	      (redder SED) or Im (bluer SED) templates rather than the Sbc
	      template.  Points with vertical lines ($\vert$) show broad
	      absorption features in their spectra which may make the estimate
	      of \mbh\ a lower limit.  Points with horizontal lines have
	      $z>2.5$.  The point that is crossed out is not included in the
	      fits (see Appendix).  The dotted lines fitted to the $z\approx2$
	      (open) points are displaced from the $z=0$ relationship by 0.3
	      ({\it a}) and 1.5 ({\it b}) mag.  The dotted circle and square
	      illustrate problematic objects which are discussed in the text
	      and in Appendix~\ref{app:problematic}.  These two objects are
	      excluded from further discussions.  In the representative
	      errorbar shown, the \mbh\ error is the scatter in the virial
	      relation, while the host galaxies have an uncertainty of roughly
	      0.3 mag.}
     \label{fig:highz}
     \vskip 0.15truein
\end{figure*}


Tables 3 and 4 summarize our estimates of the quantities needed to estimate
\mbh, as well as our final estimate of \mbh\ for those objects which have
published spectra.  For the lensed quasars we estimated the FWHM by manually
estimating the FWHM from paper copies of published spectra; two of the authors
independently estimated the widths, finding consistency.  The AGN continuum
luminosity comes from first separating the AGN from the host galaxy in the
lens image fitting as described above.  We then fit this broad-band {\it HST}
photometry ({\it V, I} and {\it H}) with a power law to estimate the continuum
luminosities entering the virial relations.  Since the virial relations depend
only on the (roughly) square root of the continuum luminosity, even
substantial errors have little effect on the estimate of
\mbh.  The corresponding values shown in Table~4 for the Ridgway et al. (2001)
and Kukula et al.  (2001) samples are from P06 after changing from the E/S0
template (P06) to the Sbc template (here) used for making $k$-corrections to
the host galaxy photometry.  For seven of the QSOs we can estimate \mbh\ from
both the C~{\sc iv} and Mg~{\sc ii}.  Four of 9 agree to better than 30\%
while the other five differ by factors of 0.5, 0.5, 1.5, 3, and 3.  The
agreement is reassuring and consistent with other studies finding that the
virial technique mass estimates have a scatter of a factor of $\sim 3$ (Kaspi
et al.  2000, Vestergaard 2002, McLure \& Jarvis 2002, Kollmeier et al.
2005).  For the 9 objects with multiple \mbh\ estimates we adopt the average.

\section {RESULTS}
\label{sect:results}

We examine the evolution of the \mbh-\lr\ relationship by dividing our host
sample into two sub-samples which span equal time intervals and comparing them
to the local relation.  Our high redshift sub-sample of 31 objects spans the
period 10--12 Gyrs ago ($1.7 \lesssim z \lesssim 4.5$) and our low redshift
sub-sample of 20 objects spans the period 8--10 Gyrs ago ($1.0\lesssim z
\lesssim 1.7$).  We use the sample of 20 nearby, normal early-type galaxies,
excluding lenticular galaxies, with \mbh\ measurements by Kormendy \& Gebhardt
(2001) and \lr\ measurements from Bettoni et al. (2003).  The \mbh--\lr\
relation for these local galaxies is well fitted by a power law
\begin {equation}
  {\rm log}\,({\mathcal M}_{\rm BH}/{\mathcal M}_\odot) = 
    -0.50 (\pm 0.06)M_R - 2.70 (\pm 1.35),
    \label{eqn:plaw}
\end {equation}
where we have converted the original Bettoni et al. (2003) result to
our standard cosmological model.  Most of the individual local objects fall
within 0.5~mag of this relation, as do nearby radio galaxies (Bettoni et al.
2003) and local quasar and Seyfert hosts (McLure \& Dunlop 2002).


\begin{figure*}
    \centerline {\includegraphics[angle=-90,width=18.0cm]{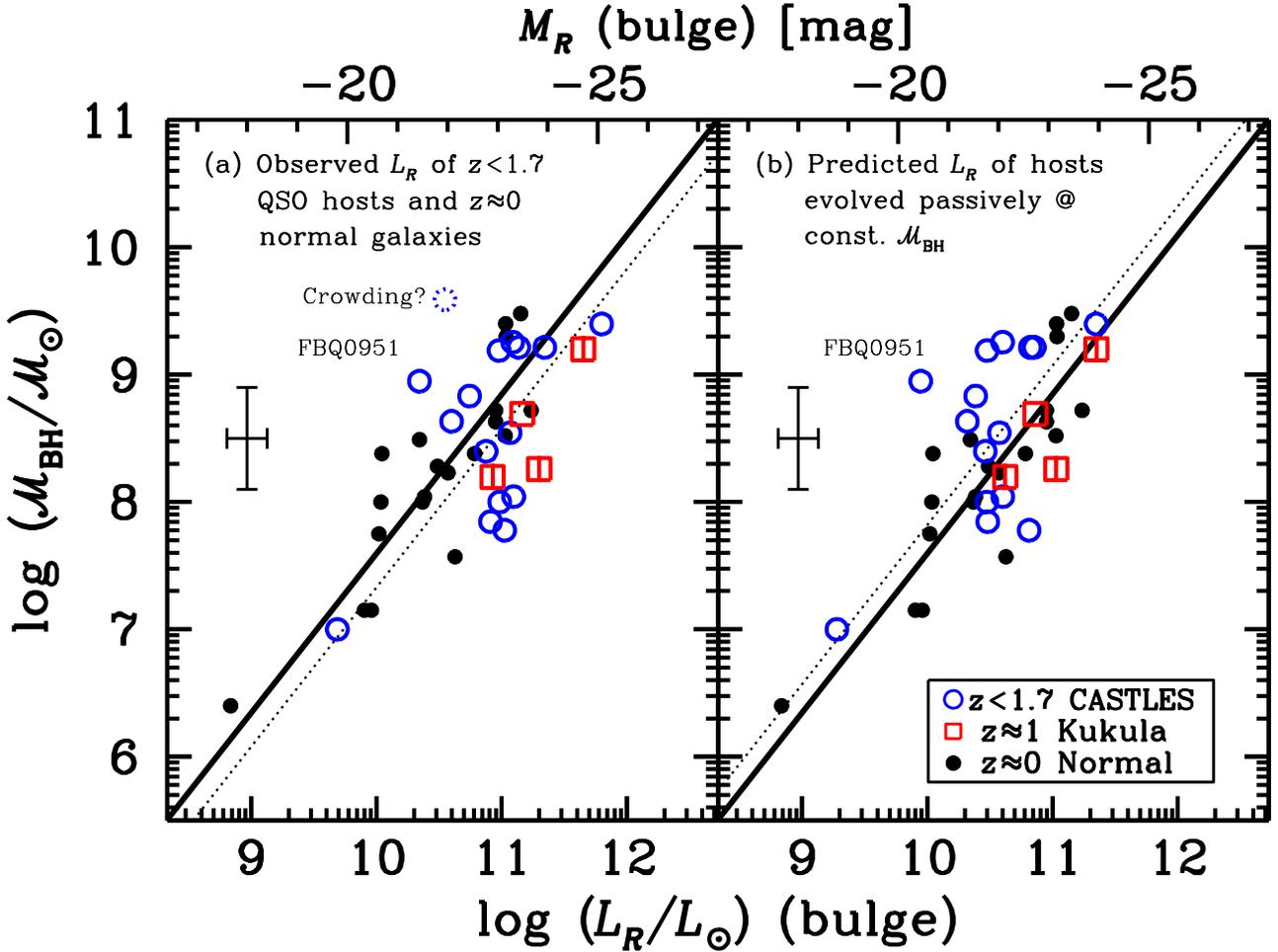}}
    \vskip -0.1truein
    \caption {The observed ({\it a}, left) and evolution corrected ({\it
	      b}, right) correlations between black hole mass and host
	      luminosity for the lower redshift ($z \lesssim 1.7$) host
	      sample. See Figure \ref{fig:highz} for details.  Points with a
	      vertical line ($\vert$) may have underestimated values of \mbh\
	      due to either broad absorption features or narrow emission line
	      components.  The dotted line is displaced from the solid line
	      representing the local \mbh-\lr\ relation by 0.5 
	      magnitude. See Appendix regarding FBQ~09051+532.}
     \label{fig:lowz}
     \vskip 0.15truein
\end{figure*}


Figs.~\ref{fig:highz} and \ref{fig:lowz} compare the high and low redshift
samples to the local sample both as observed and with evolutionary
corrections.  In addition to marking the origin of the hosts (filled circles
for the local comparison sample, open circles for the lensed hosts, triangles
for the Ridgway et al. (2001) hosts and squares for the Kukula et al. (2001)
hosts), we have also flagged objects with additional complications.  The 8
systems marked with a vertical bar ($\vert$) have absorption features near
their broad emission lines which might lead to an underestimate of the line
width and hence the black hole mass.  Objects at $z>2.5$ are marked with a
horizontal bar ($-$) because their luminosities are much more sensitive to the
choice of an SED to make the $k$-corrections.  Finally, there are four objects
discussed in the Appendix~\ref{app:problematic}, marked either by a
diagonal-cross ($\times$) or shown in dotted lines, that may be problematic.
We do not include them in the subsequent analysis.  The representative
errorbars for the gravitational lenses are also shown in the Figures, where
the uncertainty in the \mbh\ ($Y$-direction) is represented by the scatter in
the virial technique.  Whereas, for the host galaxies ($X$-direction), typical
uncertainties are roughly 0.3 magnitude for the gravitational lenses.  For
non-lenses, typical errorbars on the host luminosities are roughly 0.5-0.8 mag
(Ridgway et al.  2001, Kukula et al.  2001).

\subsection {The Evolution of the \mbh-\lr\ Relation Since $1.7 \lesssim z\lesssim 4.5$}

\label{subsect:highz}

We first consider the \mbh-\lr\ relation of the high redshift sample
($z\gtrsim 1.7$) illustrated in Fig.~\ref{fig:highz}a.  The observed relation
at these redshifts is nearly identical to the local relation -- if we fit the
high redshift sub-sample using Eqn.~\ref{eqn:plaw} with the slope fixed, then
the shift is only 0.3~mag in \lr\ or 0.2~dex in \mbh, with an RMS scatter of
0.4 dex in \mbh, and 0.8 mag in \mr\ around the best fitting line.  This
confirms the results of P06, but with three times as many data points.  The
data show that host galaxies harboring black holes of the same mass \mbh\ were
as luminous in the $R$-band at $z\gtrsim1.7$ as they are now, despite the
passage of 10~Gyrs.

Physically, we are interested in the relative growth rates of the black hole
and its host, which we can characterize by the black hole mass per unit
stellar mass, \mbh/\mstar.  This is related to the observed luminosity by the
mass-to-light ratio of the stellar population, \ml.  We will measure the
evolution of the black hole and the hosts by looking for changes in the
specific stellar mass \mbh/\mstar\ with redshift, which we can quantify by the
stellar mass deficit between the observed host and a bulge on the present day
\mbh-\mstar\ relation,
   \begin{eqnarray*} 
      \Gamma(z) & = & \frac {\mbox {\mstar}(\hbox{local bulge})} {\mbox{\mbh}} \frac {\mbox{\mbh}} {\mbox{\mstar} (\hbox{host at z})} \\
             & = & \frac {\mbox{\lr}(\hbox{local bulge})} {\mbox{\lr} (\hbox{host at z})} \times \frac {\mbox{\ml} (\hbox{local bulge})} {\mbox{\ml} (\hbox{host at z})} \\
             & = & { {\mbox{\lr}}(\hbox{local bulge}) \over {\mbox{\lr}}(\hbox{host evolved to z=0)} }
   \end {eqnarray*} 

\noindent where the last conversion holds because the luminosity of the stars in
the host will have at redshift zero is simply the observed luminosity
multiplied by the ratio of the initial and final mass-to-light ratios.  The
magnitude of the mass deficit determines the amount by which the stellar mass
of the host must grow for the system to reach the present day \mbh-\mstar\ 
relation.  We want to consider evolutionary corrections that are conservative
in the sense that any other choice will increase the stellar mass deficit.
This means that we will underestimate the required evolution in $\Gamma(z)$,
or equivalently, that we underestimate the growth in the stellar mass that
must occur in the host between its redshift and the present day due to star
formation and mergers.


For normal stellar populations, the most conservative model is to assume that
all the stars in the host were formed at some much higher redshift (we adopt
$z_f=5$) and then evolve passively to the present epoch.  If the mean
formation redshift of the stars is lower than this redshift, then the stellar
population has a lower mass-to-light ratio than in the model, so we will
overestimate the mass-to-light ratio of the observed host, underestimate the
amount by which the mass-to-light ratio will evolve, and hence underestimate
the mass deficit $\Gamma(z)$.  Adopting a higher formation redshift has little
effect since the fading rate only changes from $dM_R/dz \simeq -0.8$ mag for
$z_f=5$ to $dM_R/dz \simeq -0.6$ mag for $z_f\rightarrow \infty$ in the
Bruzual \& Charlot (2003) population synthesis models, with solar metallicity,
Salpeter IMF, and no reddening (van Dokkum \& Franx 2001).  Higher formation
redshifts will modestly reduce the stellar mass deficit, but it is a small
correction even for unphysically high mean star formation epochs.  We adopt
this passively evolving model for our standard evolutionary corrections in
estimating the stellar mass deficit and find that the mass deficit is
a factor of $\sim4$ at $z\gtrsim 1.7$.

One problem could invalidate our belief that our estimate of the mass deficit
is a lower bound, and that is the presence of significant dust extinction in
the host galaxy.  If fraction $f \leq 1$ of the stellar light is absorbed by
dust, then we have overestimated the stellar mass deficit by the same factor.
The problem, however, is that the net effect of the presence of dust on the
mass-to-light ratio is complex because higher levels of extinction are
generally associated with smaller intrinsic mass-to-light ratios due to higher
levels of star formation.  In local ULIRGs, the extra luminosity from young
stars generally prevails over dust extinction and the ULIRG has a lower
mass-to-light ratio than a passively evolving stellar population (Lonsdale,
Farrah \& Smith 2006).  For example, Arp 220 has a mass-to-light ratio of
approximately \ml$\simeq 2$ based on estimates of the stellar velocity
dispersion from Doyon et al. (2004), and this is only 50\% that of a nearby
early-type galaxy (e.g. Cappellari et al. 2005).  Thus, if Arp 220 is
representative of a typical host galaxy, the stellar mass deficit would be
$\Gamma \simeq 2$.  On the other hand, most $z\simeq 2$ galaxies have \ml$_R
\simeq 0.5$, and only the reddest galaxies reach \ml$_R \simeq 2$ whether due
to dust or age (Shapley et al. 2005).  It is difficult to determine whether
local ULIRGs are good analogies for high redshift hosts, but it is still
likely that there is a significant stellar mass deficit under the hypothesis
that the quasar hosts are dusty, star forming galaxies.

\bigskip

\noindent In summary, our overall estimate is that the stellar host for a
black hole of fixed mass at $z\gtrsim 1.7$ is on average a factor of $4$ less
massive than today if we use the Sbc model for the $k$-corrections.  Leaving
out $z\gtrsim 3$ object lowers the deficit by $\lesssim 10\%$.  The mass
deficit rises to a factor of 6 for the Im SED and falls to a factor of 3 for
the E/S0 SED, so we adopt $4^{+2}_{-1}$ as our standard estimate.  This
estimate of the deficit is designed to be a lower bound in the sense that any
other assumption about the stellar populations in the host corresponds to
adding more rapidly evolving, younger stellar populations that will increase
the correction for evolution and thus increase the mass deficit compared to a
passively evolving model.  If all the scatter were due to random effects, this
is a $6\sigma$ result, given an RMS scatter of 0.4 dex in \mbh\ and 0.8 mag
in \mr.

\subsection {The Evolution of the \mbh-\lr\ Relation Since $1 \lesssim z
\lesssim 1.7$}

When we carry out the same analysis for our low redshift sample, as shown in
Fig.~\ref{fig:lowz}, we find that the host galaxy mass deficit (Fig.~9b) is at
most a factor of two.  The observed $R$-band luminosities are about 0.5~mag
brighter than the local relation, but the passive evolution-corrected
luminosities are about 0.5~mag fainter than the local relation.
Unfortunately, we have fewer points and they exhibit more scatter (0.7 dex in
\mbh, 1.2 mag in \mr) than we found for the higher redshift sample.


\begin{figure} 
    \centerline {\includegraphics[angle=0,width=9.0cm]{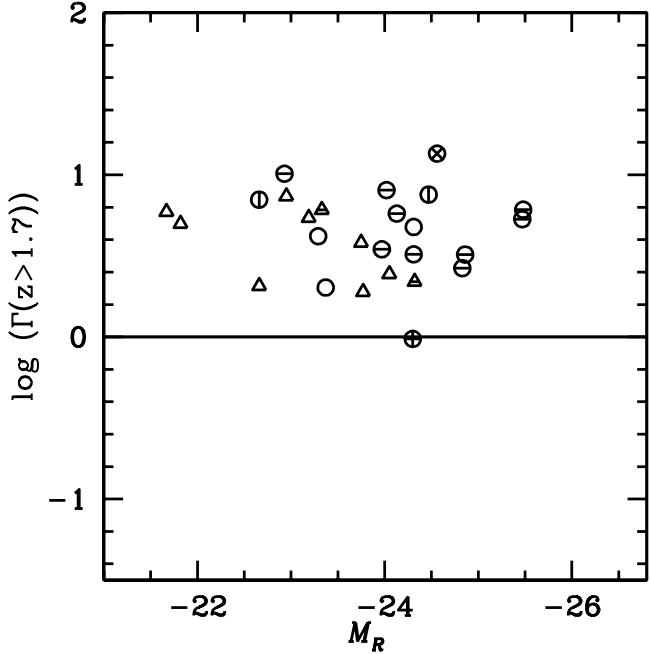}}

    \caption {The \mbh/\mstar\ ratio $\Gamma$ (relative to $z=0$) as a
	      function of rest-frame $R$-band luminosity \mr\ for galaxies
	      between look-back times of 10 and 12 Gyr ($1.7 \lesssim z
	      \lesssim 4.5$).  The circles are the lensed hosts and the
	      triangles are the directly imaged hosts from Ridgway et al.
	      (2001) and Kukula et al. (2001).  Points with $z\ge 2.5$ have a
	      horizontal line through the symbol (see Table 2), and points
	      with a vertical line may represent lower limits because \mbh\
	      may be underestimated due to narrow line contamination or broad
	      absorption features in the spectra we used to estimate \mbh.
	      Only one host, B~1422+231 (marked by $\times$) is a possible
	      non-detection, so Malmquist bias is unimportant.  }
    \label{fig:evolvel}
    \vskip 0.15truein
\end{figure}


The larger scatter is due to several factors.  First, we have fewer objects in
this redshift bin and several of them may be problematic (see the
Appendix~\ref{app:problematic}).  One lensed host, SBS~0909+532 ($z=1.38$) we
discard entirely because the fits are unstable.  Another lensed host,
FBQ~0951+2635 ($z=1.24$) is well-detected and modeled, but lies 3 mag from the
general trend.  We include it in the fits since we lack an objective basis for
rejecting it, but it does bias our fits towards fainter hosts by 0.15 mag and
accounts for a third of the implied evolution of this redshift bin.  Six of
the unlensed hosts may have biased estimates of the black hole mass due to
beaming effects, contributions from narrow emission lines or broad absorption
lines (see Appendix~\ref{app:assumptions}).  As a second source of scatter,
the stellar populations in this redshift range may be more heterogeneous simply
because the universe is older. In our high redshift bin, all stars must be
$<3$~Gyr old (at $z=2$), while in this bin they can be up to $6$~Gyr old.  An
increasing diversity in the average age of the stellar populations will appear
as scatter.  A third point is that the hosts seem to make up much of the
stellar mass deficit we observed for the high redshift hosts in this lower
redshift range, leading to a systematic offset between the high- and
low-redshift object ends of the $1\lesssim z\lesssim1.7$ range.

Thus, we conclude that the \mbh-\lr\ relation also existed 8--10~Gyr ago, but
with a significantly smaller offset from the local relation than we observed
in the higher redshift sample, and with a larger RMS scatter (0.7 dex in \mbh,
1.2 mag in \mr).  Unfortunately, with the small numbers of objects and the
increased scatter among them, it is difficult to quantify this qualitative
trend precisely.  At most, a stellar mass deficit of a factor of two is
permitted, but the statistical significance appears weak ($\sim 1\sigma$).
Thus the host galaxies could be consistent with being passively evolving since
$z\sim 1$.

\section {DISCUSSION AND CONCLUSION}
\label{sect:conclusion}

\begin{figure*} 
    \centerline {\includegraphics[angle=0,width=13.0cm]{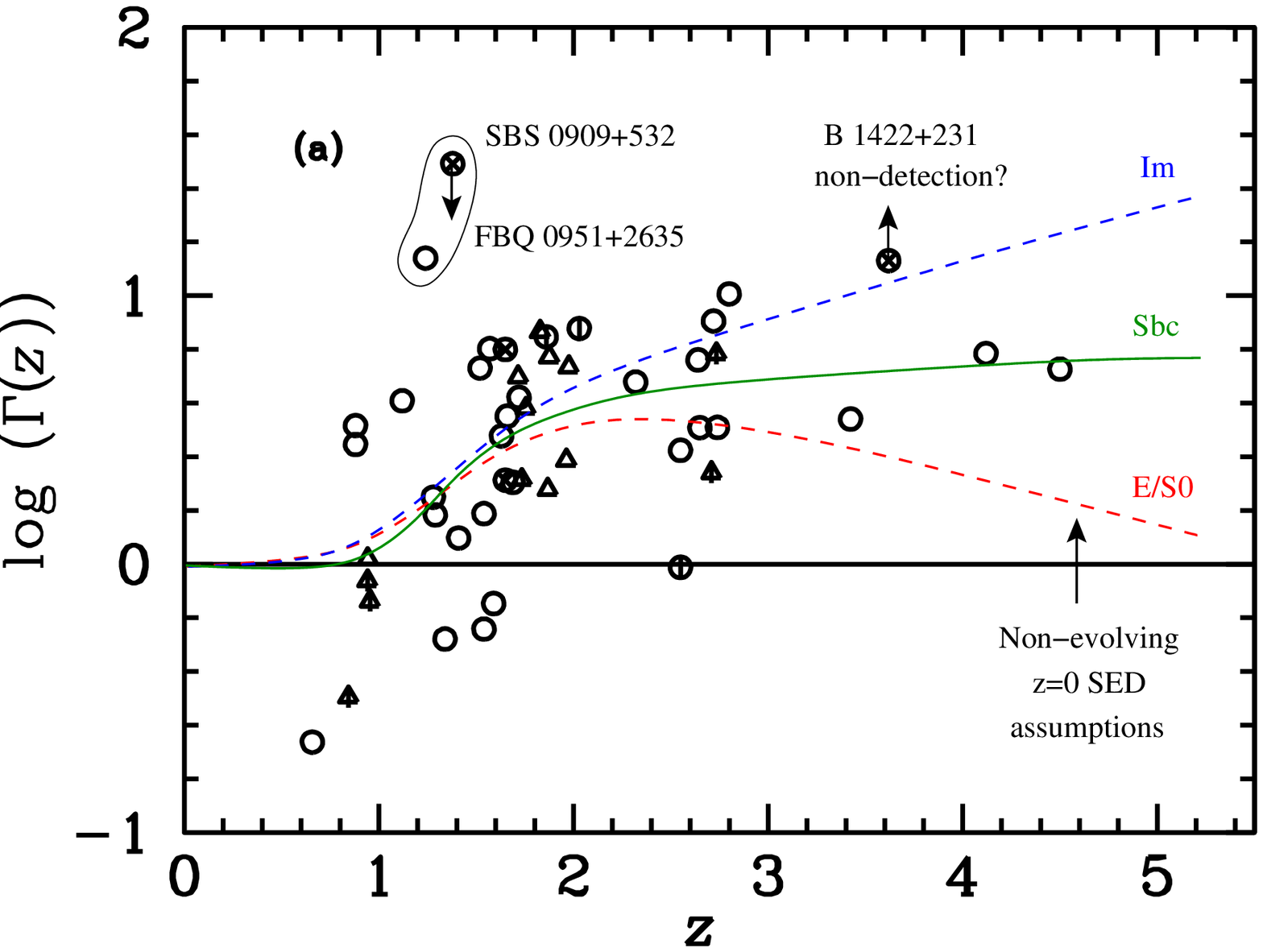}}
    \centerline	{\includegraphics[angle=0,width=13.0cm]{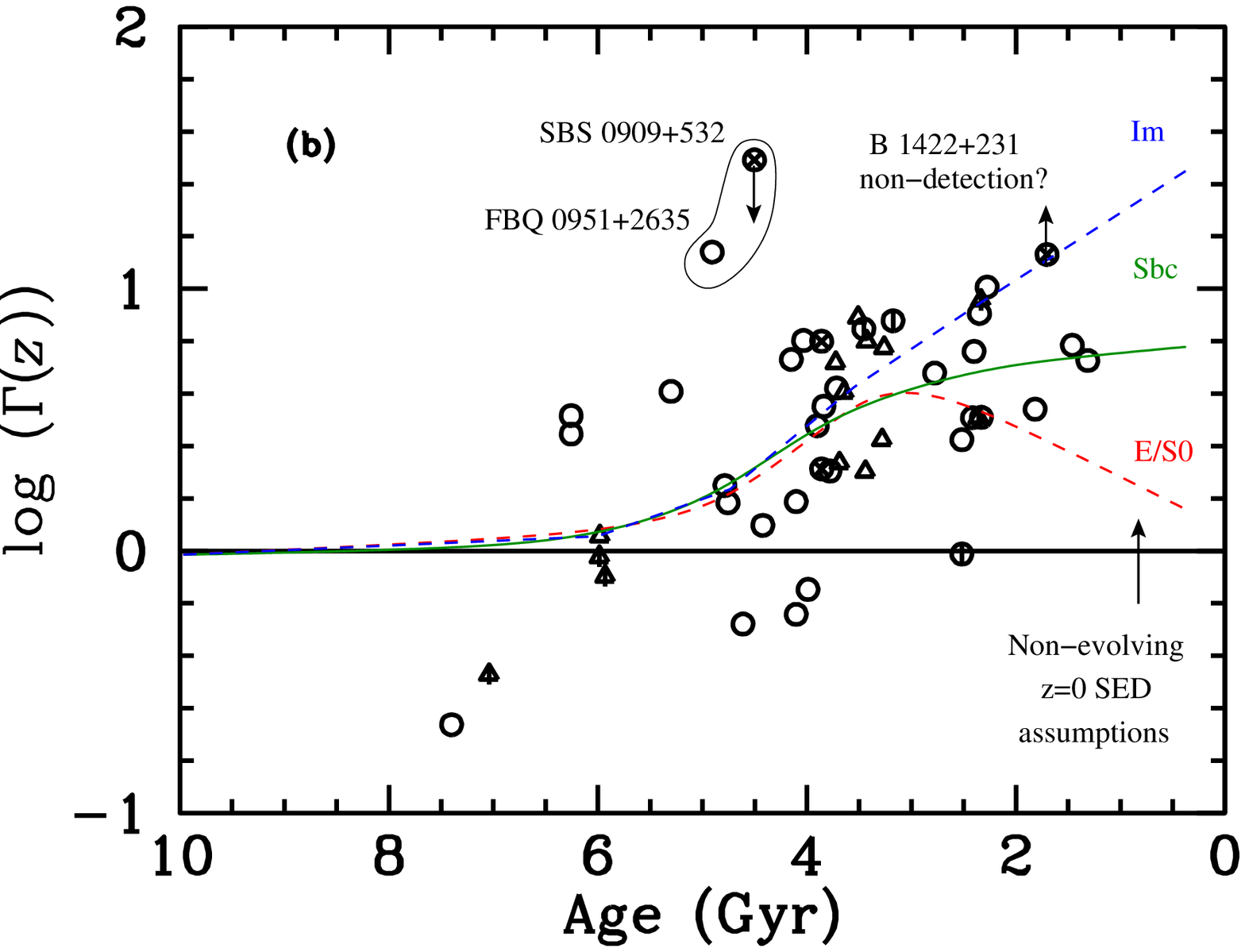}}

    \caption {The \mbh/\mstar\ ratio $\Gamma$ normalized to $z=0$, or
	      equivalently the host mass deficit, as a function of redshift
	      (Panel~{\it a}) and time (Panel~{\it b}).  The symbols are the
	      same as in Figure \ref{fig:evolvel}.  The curves roughly
	      indicate the consequences of using different non-evolving,
	      $z=0$, SED templates to compute the $k$-corrections.  All data
	      points are based on assuming the Sbc SED (solid line) as
	      in Fig.~\ref{fig:highz}.  All the scenarios assume the same
	      passive-evolution corrections.  Changing to any more rapidly
	      evolving model would increase the \mbh/\mstar(z) ratio.}
    \label{fig:evolvet}
    \vskip 0.15truein
\end{figure*}


Figs.~\ref{fig:evolvel} and \ref{fig:evolvet} summarize the results by showing
the stellar mass deficit $\Gamma (z)$ of the hosts as a function of luminosity
and redshift.  Figure~\ref{fig:evolvet} is equivalent to the evolution of the
\mbh/\mstar\ ratio.  While we see little correlation of $\Gamma(z)$ with
luminosity (Fig.~\ref{fig:evolvel}), we see considerable evidence for a
correlation with redshift and look-back time (Fig.~\ref{fig:evolvet}).  A
detailed study of the evolution with redshift is now limited by small number
statistics at $z\gtrsim 3$ and $z\lesssim 1.5$.  Crudely it appears that
$\Gamma(z)$ evolves little to $z \sim 1$, rises to $\Gamma(z)\sim 4$ at $z
\sim 2$ and then saturates at a factor of $\approx 6$.  If we assumed a bluer
SED for the $z>1.7$ sample, the rising trend would continue further, while if
we assumed a redder SED, it would decline at the higher redshifts
(Fig.~\ref{fig:evolvet}).  We will address this issue in more detail as we
acquire color information for the hosts.

If real, this trend must arise from a combination of ongoing star formation
and mergers between galaxies.  Star formation provides a very simple
explanation, particularly since the evolution of $\Gamma(z)$ looks remarkably
similar to estimates of the star formation history in which the rates are
roughly constant to $z \sim 1$ and then decline rapidly, with at least 50\% of
the stellar mass in place by redshift unity (e.g. Bauer et al. 2005 or Feulner
et al. 2005).  

A simple way to parameterize the amount of star formation needed to achieve
agreement with normal galaxies today is through estimating the required amount
of specific star formation, $\phi$, the SFR per unit mass.  Thus, a quasar
host with mass deficit $\Gamma$ would require an average star formation rate
of $\left<\phi\right>~=~(\Gamma~-~1)/\Delta t \mbox{ [Gyr}^{-1}]$, if all the
missing stellar mass forms in the time span of $\Delta t$.  Our estimated mass
deficit of $\Gamma=4$ between $z=2$ and $z=1$ ($\Delta t=2.5$ Gyrs) leads to
$\left<\phi\right> = 1.2$~Gyr$^{-1}$.  Such a star formation rate, though high
for the most massive objects in our study, is in fact observed in $z\sim 2$
galaxies (Reddy et al.  2006), which would also require quasar hosts at $z\sim
2$ to be very actively star forming.  However, if not all the mass deficit is
fully accounted for by $z\approx 1$, as possibly our data suggest, then the
star formation rate needed can be as low as $\left<\phi\right> =
0.6$~Gyr$^{-1}$, which is seen in some distant red galaxies (Reddy et al.
2006).  The typical star formation in a Lyman break galaxy of $\phi \gtrsim 2$
(Reddy et al. 2006; Ouchi et al. 2004; Reddy, Steidel, \& Erb 2005) is more
than sufficient to eliminate the stellar mass deficit of the bulge hosting a
$10^8$\msol\ BH.  By adding a modest amount of internal extinction or
extending the period of star formation, the required SFR could be even lower.
The rest-frame UV colors of hosts at high redshifts tend to be fairly blue and
are consistent with on-going star formation (Jahnke et al.  2005).  On the
other hand, if the central BHs gain significantly in mass over this time
through accretion, as opposed to BH merging, then the required star formation
rate would have to rise proportionally.

Many massive galaxies will also experience major mergers over this redshift
range, which can also lead to evolution in $\Gamma(z)$.  Mergers of pure bulge
galaxies can only move systems along the \mbh-\lr\ relation, but major mergers
of systems with disks increase $\Gamma(z)$ as the disks are disrupted and
added to the bulge. A merger of two identical galaxies with a bulge-to-disk
ratio of $(B/D)$ reduces $\Gamma(z)$ by a factor of $1+(B/D)^{-1}$, if all the
disk stars are incorporated into the bulge.  The process can reduce
$\Gamma(z)$ by large factors only if the hosts are disk-dominated systems,
which currently cannot be constrained given our relatively shallow exposures.
Detailed studies of the highest signal-to-noise objects will be discussed in
Peng et al.  (2006 in preparation).  Mergers with smaller satellites would
contribute to decreasing $\Gamma(z)$ but not by large factors.  Thus, mergers
probably contribute to the downward evolution of $\Gamma(z)$ but may not
dominate the evolution. On the other hand, it has been suggested that if the
BH masses were to grow by $\gtrsim 10\times$ since $z\approx 2$, {\it and} the
local \mbh-\lbulge\ relation were to greatly {\it steepen} for \mbh$\gtrsim
3\times10^9$\msol, the amount of galaxy mass growth required can be less than
our inference.  However, this appears not to be an issue, given a recent work
which extends the local \mbh-\lbulge\ relation to some of the largest BH
masses (Lauer et al. 2006, in preparation).

Improving our results depends on obtaining three new types of data.  First, we
need to quantify the color and structure of the host galaxies to better
understand their structures and stellar populations.  In our structural study
of the hosts (Peng et al. 2006, in preparation), we find that the typical
hosts have effective radii of $r_e \lesssim 3$-5 kpc that are significantly
smaller than present-day galaxies with the same \mbh, and that they may be
less concentrated.  As mentioned earlier, the hosts seem to have fairly blue
rest-frame UV colors that imply star formation rates of order $\sim
20$~\msol~yr$^{-1}$ (Jahnke et al. 2005).  Second, we need to increase the
number of hosts with measured properties, particularly in the redshift range
$0.5 < z < 2.0$, to better probe the evolution of $\Gamma(z)$ and to begin
studying the evolution of the dispersion of the \mbh-\lr\ relation with
redshift.  Third, it would be a very useful check of the virial relations if
the sizes of the broad line emitting regions of quasars at $z>1$ could be
measured directly as a check for evolution in the relations.  While
reverberation mapping methods probably cannot be used due to the time scales,
it is likely that gravitational lensing can be used to make the measurements
by examining the microlensing of the broad line region by stars in the lens
galaxy.  There is considerable evidence now that the continuum and emission
line regions are differentially microlensed (e.g. Wayth, O'Dowd
\& Webster 2005; Popovi\'c et al. 2006), and the next challenge is to use
these differences to estimate the sizes of the regions (Richards et al. 2004;
Keeton et al. 2005).  This has been done for the continuum regions of several
lensed quasars, the $z=1.7$ quasar Q~2237$+$0305 in particular (see Kochanek,
Schneider, Wambsganss 2004 and references therein).

\bigskip
\bigskip

\noindent We thank Sandy Faber, Marianne Vestergaard and Xiaohui Fan, Tod
Lauer, Luis Ho, Jenny Greene, Mike Santos, and Roeland van der Marel for
enlightening discussions on various issues of the black hole/galaxy
correlation, and John Moustakas for discussions on $k$-correction issues.  We
also thank the anonymous referee for a thoughtful report.  The work of CYP was
performed in part under contract with the Jet Propulsion Laboratory (JPL)
funded by NASA through the Michelson Fellowship Program.  JPL is managed for
NASA by the California Institute of Technology.  CYP is also grateful to STScI
for support through the Institute Fellowship Program.  We also thank the SDSS
collaboration for providing such a valuable database.  Support for the CASTLES
project was provided by NASA through grant numbers GO-7338, GO-7495, GO-7887,
GO-8175, GO-8252, GO-8804, GO-9133, GO-9138, GO-9375, and GO-9744, from the
Space Telescope Science Institute, which is operated by the Association of
Universities for Research in Astronomy, Inc., for NASA, under contract
NAS5-26555.  Our research was also supported by the Smithsonian Institution.
This research has made use of the NASA/IPAC Extragalactic Database (NED) which
is operated by the Jet Propulsion Laboratory, California Institute of
Technology, under contract with the National Aeronautics and Space
Administration.

\begin{appendix}

\section {SYSTEMATIC ERRORS DUE TO MALMQUIST BIAS AND ASTROPHYSICAL
ASSUMPTIONS}

\label{app:assumptions}

Peng et al. (2005) discussed the known sources of systematic and random
uncertainties in the \mbh-\lr\ and the \mbh-\mstar\ relations in detail, but
it is worth reviewing the major concerns.  We have already discussed our
conservative approach to correcting and evolving the SED of the host galaxy
and a dusty starburst scenario (\S~\ref{subsect:kcorr}), where by conservative
we mean that alternative choices would lead to larger estimates of the
evolution in \mbh/\mstar\ than we report.  Other issues that could affect the
results are Malmquist biases, PSF model problems, and the applicability of the
local virial relations to high redshift systems.

Malmquist bias, where we detect only the most luminous hosts, can matter little
because our sample is nearly complete. There is only one system out of 51 for
which we are uncertain about the host detection.  Moreover, the effect of
missing lower luminosity hosts would only increase the necessary amount
of evolution because it would mean that the actual mean host luminosity is
fainter than our estimates.  Reducing our estimate of the evolution in 
\mbh/\mstar\ means that we must be {\it under-estimating} the host galaxy
luminosities by 1.5~mag, {\it over-estimating} the black hole masses by
0.6~dex (a factor of 4), or some combination of both effects.  It is possible
for the host luminosities to be underestimated by 0.2-0.3 magnitudes due to
either problems in the PSF models (S\'anchez et al.  2004) or limits to our
surface brightness sensitivity, but this effect is much smaller for lensed
than for unlensed hosts.

P06 demonstrated that the systematic errors due to the PSF model for the
Kukula et al. (2001) and Ridgway et al. (2001) samples are small compared to
the observed evolutionary trends.  In Appendix~\ref{app:complexity} we present
results for comparing the fits to long and short exposures of the same lensed
object, also finding no significant biases.  Moreover, our sample of lensed
hosts is less sensitive to this problem than traditional samples of high
redshift host galaxies.  This reduction in systematic uncertainties is not
offset by the need to model the lensing effects.  As we discuss in the
Appendices~\ref{app:degeneracies}, the lens model introduces no significant
biases unless we assume that galaxies have no dark matter halos -- an
assumption leading to a host of other cosmological complications.  Finally, we
note that the agreement of the estimated properties of lensed and unlensed
host galaxies suggests that neither has major systematic problems.

The conclusions in this study are sensitive to systematic shifts between the
local virial calibrations for \mbh\ and higher redshifts.  There is no
theoretical mechanism for such changes because the virial relation should
depend only on the local accretion physics of black holes.  Netzer (2003),
Vestergaard (2004) and Baskin \& Laor (2005) discuss why they should be
applicable to luminous quasars and the potential caveats.  In fact, Baskin \&
Laor (2005) found that for C~{\sc iv} lines wider than than 4000 km~s$^{-1}$,
as for most objects here, the virial BH mass estimator may actually be biased
towards {\it low} \mbh\ estimates, and such a correction would increase the
estimated evolution.  Internal to our sample we see no systematic offsets in
the \mbh\ estimates from the Mg~{\sc ii} and C~{\sc iv} for the 6 objects with
both lines.  A much larger study by Kollmeier et al.  (2005) also finds
consistent results from different lines, and considerable evidence that there
can be little systematic scatter in the \mbh\ estimates for quasars of
comparable redshift and luminosity.

Lastly, as discussed in Section~\ref{subsect:kcorr}, a dusty starburst of the
type like Arp~220 has a bluer SED and a lower \ml\ ratio than a local stellar
bulge.  Moreover, even without the compensating effects of a reduced \ml\
ratio, and the fact that dust is included {\it implicitly} in our choice of
SED templates, it would take an additional 1.5~mag of $R$-band extinction to
weaken our conclusions.  This level of {\it absolute} extinction is more
typical of the extinction for Lyman break galaxy in the restframe ultraviolet
(1600~\AA, Shapley et al.  2001) than our restframe optical measurements
($H$-band is still only 2900~\AA\ in the rest frame of our highest redshift
host, at $z=4.5$).  In summary, the net effect of a dusty star burst having a
lower \ml\ ratio than our SED assumptions would imply a {\it larger} amount of
evolution in the \mbh/\mstar\ ratio than redder, passively evolving, host
galaxies.

\section {LENS MODEL DEGENERACIES}

\label{app:degeneracies}

The complexities of the lens modeling process coupled with image fitting are
often causes of concern to non-practitioners even if they are well understood
by the practitioners. Here we briefly discuss some of the most common concerns.  
Readers interested in a detailed review of lens models and
their limitations should see Kochanek, Schneider, \& Wambsganss (2004).

Gravitational lenses are extraordinarily good at constraining the angular
structure of the gravitational potential, particularly when the host galaxy
forms a well-defined Einstein ring around the lens (e.g.
Fig.~\ref{fig:pg1115}).  Analyses of these rings show that the gravitational
potentials are centered on the luminous lens galaxies and that the deviations
of the potential from that of an ellipsoid are consistent with zero (Yoo et
al. 2005).  This means that the only real freedom in our model is adjusting
the radial density profile of the lens, which corresponds to adjusting the
mass of the dark matter halo of the lens relative to the visible galaxy.

For many lensing galaxies, the observations cannot determine the exact radial
density profile because of the mass sheet degeneracy.  An example of this
degeneracy is the fact that most of our lensing galaxies reside inside a weak
galaxy cluster which acts like a uniform mass sheet at the scale of the lensed
images that causes an additional, isotropic, magnification.  If we take a lens
and add a constant surface density sheet, $\kappa$ in units of the critical
density for lensing, then the only observable quantity that is altered is the
time delay between the images.  The effect on the properties of the host are a
simple magnification -- if additional surface density is added, then the scale
length of the host shrinks by factor of ($1-\kappa$) and its flux drops by a
factor of $(1-\kappa)^2$.  The effect of changing the density profile of the
lensing galaxy itself is a local analog of the mass sheet degeneracy.  Near
the Einstein ring where we observe the hosts, the surface density of the SIE
is $\kappa \simeq 0.5$ while that of a typical constant mass-to-light ratio
\ml\ model for the lens would be $\kappa \simeq 0.2$ because the SIE mass
distribution is more centrally concentrated.  The SIE model magnifies hosts
more than the constant \ml\ model, so if galaxies lacked dark matter we would
underestimate their intrinsic fluxes by roughly $(1-0.2)^2/(1-0.5)^2\simeq
2.5$ or 1 magnitude.  In practice, we know from studies of individual lenses
(e.g. Impey et al.  1998; Le\'har et al. 2000; Mu\~noz, Kochanek, \& Keeton
2001), statistical properties of lens samples (e.g. Chae et al. 2002, Zhang
2004, Chen 2005, Rusin \& Kochanek 2005), and dynamical studies (Rix et al.
1997, Treu \& Koopmans 2004, Treu et al. 2006), that the typical lens has a
radial density distribution that is reasonably well modeled by an SIE in the
region where we observe the host galaxies.  There is likely to be a scatter
about this typical model, but it is probably at a level of 0.1--0.2~mag that
does not represent a significant contribution to our uncertainties given the
other sources of error.  In our comparisons with the unlensed hosts of Kukula
et al.  (2001) and Ridgway et al.  (2001) we find that the properties of the
lensed and unlensed hosts are mutually consistent, adding empirical
confirmation to this assertion.

\section {UNCERTAINTIES DUE TO MODELING COMPLEXITY}

\label{app:complexity}

The lens modeling and the image fitting process required to simultaneously
model everything seen in each image are a non-trivial challenge.  The
complexity of the models has some effect on our inferences about the hosts,
but little effect on our conclusions, by examining the lenses shown in
Figures~\ref{fig:he1104}--\ref{fig:bri0952}.  Remember that shifts in the host
luminosity larger than $\sim 1$~mag are required to significantly affect
our discussion.

Figures \ref{fig:he1104}--\ref{fig:rxj0911} show deep (5 orbit) images of
HE~1104$-$1805 ($z_s = 2.32$), PG~1115$+$080 ($z_s=1.72$), RXJ~0911+0551
($z_s=2.8$), and B~1600+434 ($z_s=1.59$), while Figure~\ref{fig:bri0952} is
the shallower (2 orbit) image of BRI~0952$-$0115 ($z=4.5$).  In each case, the
host is readily visible after subtracting the quasar and has a spatial
structure virtually impossible to mimic with PSF errors or a mismodeled lens
galaxy.  Because of our restriction on the diameter of the Einstein ring
($>0\farcs8$), the quasar images and the lens galaxy are well-separated.
Thus, even though we must use a 7-parameter model for the lens galaxy light
profile, these parameters have little covariance with those of the host.  For
the less well-constrained lenses, two-image lenses with less well-detected
hosts, there can be degeneracies in the lens model trading off the elliptical
structure of the lens and the external shear.  But these are degeneracies that
must hold the pattern of ray deflections fixed relative to the data, so they
have negligible effects on the inferred structure of the host.  Empirically,
the covariances between the parameters for the lens galaxy and the lens model
negligibly broaden the uncertainties in the host properties.

One of our primary concerns is inferring the host parameters in shallow images
where the outer regions of the host may be poorly detected.  We can test for
problems by comparing the host properties derived from deep 5-orbit images
shown in Figs.~\ref{fig:he1104}--\ref{fig:rxj0911}, to those derived from
earlier single orbit images.  Both the depth of the observations and the PSF
have changed.  For HE~1104--1805 and PG~1115+080 the host luminosities found
from the deep images are 0--0.15~mag brighter and there are 10-20\% changes in
the estimated effective radii.  Such changes are quite comparable to the
effects we find from using different model PSFs.  Where the host is
significantly fainter, as in RXJ~0911+0511, the upper limit on the luminosity
estimate from the shallow image is 0.15~mag brighter than in the deep image.
B~1600+434 is our most extreme example of the lens and host emission
overlapping.  We also require a two-component bulge+disk model for the lens
and need to mask the dust lane in the disk of the lens galaxy.  The difference
in host luminosity between the deep and shallow images is at worst 0.25~mag,
caused partly by uncertainties in the bulge-to-disk ratio of the lens galaxy
(it ranges from 0.5 to 2), and partly by the lens model.  For instance, if we
artificially change the lens bulge size by 50\% in the deep image, our
estimate of the host luminosity changes by 0.3~mag.  We also find that the
uncertainties in the structures of the lensed hosts are markedly less than
those for the unlensed hosts.  Presumably this is a combination of better
resolving the hosts because of the magnification and distorting it into a
shape that has significantly less correlation with the structure of the PSF.
In short, modeling lensed hosts may appear fraught with uncertainties, but in
practice we can estimate their parameters at least as well as for unlensed
hosts and in some cases markedly better.

\section {PROBLEMATIC OBJECTS}

\label{app:problematic}

Several objects are problematic (marked by embedded $\times$ symbol or a
dotted circle/square in Figures \ref{fig:highz} and \ref{fig:lowz}).  Here we
discuss the reasons for the outliers and why we have excluded some objects
from our sample.

\bigskip

\noindent {\it 4C~45.51} -- this unlensed QSO from Kukula et al. (2001) is an
extremely powerful radio source ($\sim 3\times10^{28}$ to $\sim10^{29} \mbox{
W Hz}^{-1}$ from 151 MHz to 37 GHz, e.g. Wiren et al.  1992; Hales, Baldwin \&
Warner 1993) which implies such a high Eddington ratio ($\epsilon_{Edd}
\approx 0.7$) that the source is probably beamed towards us.  If beamed
emission contributes to the optical continuum, then the BH mass estimate is
suspect (see also P06).

\bigskip

\noindent {\it MG~2016+112} -- this lensed quasar is peculiar on two grounds.
First, it is a narrow line quasar (NLQ), and local studies of NLQ (narrow-line
Seyfert 1) find that many are radiating near the Eddington limit and may be
beamed (e.g. Greene \& Ho 2004, Bian \& Zhao 2004, Botte et al. 2005).
Indeed, Botte et al. (2005) find larger BH masses using the \msig\ technique
than through the virial estimate, casting significant doubt on the reliability
of BH mass estimates in NLQs.  Second, the extended optical emission observed
in MG~2016+112 is also associated with lensed X-ray emission (Chartas et al.
2001) and images of the radio jets (Koopmans et al.  2002), suggesting that
some of the optical emission is related to extended emission from a jet rather
than from the host galaxy.

\bigskip

\noindent {\it RXJ~0921+4528} -- Based on our models of the host galaxies, we
have concluded that this system (shown as a dotted circle in
Figure~\ref{fig:highz}) is a binary quasar rather than a gravitational lens.  This
was already a concern of Mu\~noz et al. (2001); in our analysis we see no
signs of an Einstein ring structure to the host galaxies.  Treated as a lens,
the host galaxy lies far from the general trends with an inferred mass deficit
of $\Gamma(z) \approx 30$, while treated as a binary it has a mass deficit of
$\Gamma(z) \approx 4$ that is typical of the other host galaxies at its
redshift.

\bigskip

\noindent {\it B~1422+231} -- this lensed QSO is excluded because we only have
an upper limit on the luminosity of the host.  Viewed as a marginal detection,
the host luminosity is consistent with the general correlations, but we do
not include it in the analysis.

\bigskip

\noindent {\it SBS~0909+532} -- the host galaxy images in this system merge
into the emission from the lens galaxy and we lack a PSF model that works well
for these observations.  Thus we probably underestimate the host luminosity.
We do not include it in the analysis.

\bigskip

\noindent {\it FBQ~0951+2635} -- as with B~1600+434, the lens is an edge on,
late-type spiral galaxy that must be modeled using both a bulge and a disk.
Unfortunately, the system is also more compact than B~1600+434, so the
photometric models are less stable.  Problems with the photometric model are
real but seem to be too small to explain the offset.  For example, using a
bulgeless model for the lens galaxy only increases the host luminosity by
0.4~mag, which still leaves the system 2.6 mag off the general trend.
Similarly, \mbh\ may be underestimated because the Mg~{\sc {ii}} line used to
estimate it is heavily contaminated by Fe line emission, but the magnitude of
any potential error is too small to explain the offset.

\end {appendix}

\begin {references}



\reference{} 
Aretxaga, I., Terlevich, R. J., \& Boyle, B. J. 1998, \mnras, 296, 643


\reference{}
Bade, N., Siebert, J., Lopez, S., Voges, W., \& Reimers, D. 1997, \aap, 317, L13


\reference{} 
Barth, A.~J., Greene, J. E., \& Ho, L.~C. 2005, \apj, 619, L151

\reference {}
Baskin, A., \& Laor, A. 2005, \mnras, 356, 1029

\reference {}
Bauer, A.~E., Drory, E., Hill, G.~J., \& Feulner, G. 2005, \apj, 621, 89


\reference {}
Bettoni, D., Falomo, R., Fasano, G., \& Govoni, F. 2003, \aap, 399, 869

\reference {}
Bian, W., \& Zhao, Y. 2004, \mnras, 352, 823

\reference {}
Blandford, R. D., \& McKee, C. F. 1982, \apj, 255, 419

\reference {}
Boroson, T. A. 2003, \apj, 585, 647

\reference {}
Borys, C., Smail, I., Chapman, S.~C., Blain, A.~W., Alexander, D.~M., \&
Ivison, R.~J., \apj, 635, 853

\reference {}
Botte, V., Ciroi, S., di Mille, F., Rafanelli, P., \& Romano, A. 2005, \mnras,
356, 789



\reference{}
Bruzual, G., \& Charlot, S. 2003, \mnras, 344, 1000

\reference{}
Cappellari,  M., et al. 2005, astro-ph/0505042


\reference{}
Chae, K.-H., et al.\ 2002, Physical Review Letters, 89, 151301 

\reference{}
Chartas, G. Bautz, M., Garmire, G., Jones, C, \& Schneider, D.P., 2001, ApJL, 550, 163

\reference{}
Chavushyan, V.~H., Vlasyuk, V.~V., Stepanian, J.~A., \& Erastova, L.~K. 1997, \aap, 318, 67

\reference{}
Chen, D.-M. 2005, \apj, 629, 23

\reference{}
Claeskens, J.-F., Surdej, J., \& Remy, M. 1996, \aap, 305, L9

\reference{}
Coleman, G. D., Wu, C.-C., \& Weedman, D. W. 1980, \apjs, 43, 393


\reference{}
Cox, P., et al. 2002, \aap, 387, 406

\reference{}
Devriendt, J. E. G., Guiderdoni, B., \& Sadat, R. 1999, \aap, 350, 381

\reference{}
Di Matteo, T., Springel, V., \& Hernquist, L. 2005, Nature, 433, 604

\reference{}
Doyon, R., et al. 1994, \apjl, 437, 23

\reference{}
Dunlop, J. S., McLure, R. J., Kukula, M. J., Baum, S. A., O'Dea, C. P., \&
Hughes, D. H. 2003, \mnras, 340, 1095


\reference{}
Erwin, P., Graham, A.~W., \& Caon, N. 2002, astro-ph/0212335

\reference{}
Falco, E. E., et al. 1999, \apj, 523, 617

\reference{}
Falomo, R., Kotilainen, J., \& Treves, A. 2001, \apj, 547, 124

\reference{}
Fassnacht, C.~D., \& Cohen, J. 1998, \aj, 115, 377

\reference{}
Fassnacht, C.~D., et al. 1999, \aj, 117, 658

\reference{}
Ferrarese, L., \& Merritt, D. 2000, \apjl, 539, L9

\reference{}
Ferrarese, L., Pogge, R. W., Peterson, B. M., Merritt, D., Wandel, A., \&
Joseph, C. L. 2001, \apj, 555, 79

\reference{}
Feulner, G., Goranova, Y., Drory, N., Hopp, U., \& Bender, R. 2005, \mnras,
358, 1


\reference{} 
Gebhardt, K., et al. 2000a, \apj, 539, L13

\reference{} 
------. 2000b, \apj, 543, L5

\reference{}
Gehren, T., Fried, J., Wehinger, P. A., \& Wyckoff, S. 1984, \apj, 278, 11

\reference{} 
G\'omez-Alvarez, P. 2005, Conference on ``25 Years After Discovery:
Some Current Topics on Lensed QSOs", Ed. L.~J. Goicoechea (UC, Spain), p. 23

\reference{} 
Graham, A. W., Erwin, P., Caon, N., \& Trujillo, I. 2001, \apjl, 563, 11

\reference{} 
Greene, J.~E., \& Ho, L.~C. 2004, \apj 610, 722

\reference{} 
Greene, J.~E., \& Ho, L.~C. 2005, astro-ph/0512461




\reference{}
Hammer, F., Rigaut, F., Angonin-Willaime, M.-C., \& Vanderriest, C. 1995, \aap, 298, 737

\reference{}
H\"aring, N., \& Rix, H.-W. 2004, \apjl, 604, 89

\reference{}
Hales, S. E. G., Baldwin, J. E., \& Warner, P. J., 2003, \mnras, 263, 25

\reference{}
Hewett, P.~C., Irwin, M.~J., Foltz, C.~B., Harding, M.~E., Corrigan, R.~T., Webster, R.~L., \& Dinshaw, N., 1994, \aj, 108, 1534

\reference{}
Ho, L.~C. 1999, in Observational Evidence for Black Holes in the Universe,
ed. S. K. Chakrabarti (Dordrecht: Kluwer), 157

\reference{}
Ho, L.~C. 2002, \apj, 564, 120

\reference{}
Ho, L.~C. 2005, \apj, 629, 680

\reference{} 
Hogg, D. W., Baldry, I. K., Blanton, M. R., \& Eisenstein, D. J.  2004,
astro-ph/0210394

\reference{}
Hutchings, J.~B., Crampton, D., \& Campbell, B. 1984, \apj, 280, 41

\reference{}
Hutchings, J.~B., Frenette, D., Hanisch, R., Mo, J., Dumont, P.~J., Redding,
D.~C., \& Neff, S.~G. 2002, \aj, 123, 2936

\reference{}
Hutchings, J. B. 2003, \aj, 125, 1053

\reference{}
Impey, C. D., Falco, E. E., Kochanek, C.~S., Leh\'ar, J., McLeod, B.~A., Rix,
H.-W., Peng, C.~Y., \& Keeton, C.~R. 1998, \apj, 509, 551

\reference{}
Inada, N. et al. 2003, \aj, 126, 666

\reference{} 
Isaak, K. G., Priddey, R. S., McMahon, R. G., Beelen, A., Peroux, C., Sharp, 
R. G., \& Withington, S. 2002, \mnras, 329, 149

\reference{} 
Ivanov, V. D., \& Alonso-Herrero, A. 2003, Ap\&SS, 284, 565


\reference{} 
Jahnke, K. et al. 2004, \apj, 614, 568


\reference{}
Kaspi, S., Smith, P.~S., Netzer, H., Maoz, D., Jannuzi, B.~T., \& Giveon,
U. 2000, ApJ, 533, 631

\reference{} 
Kaspi, S., Maoz, D., Netzer, H., Peterson, B. M., Vestergaard, \&
M., Jannuzi, B. T. 2005, \apj, 629, 61


\reference{} 
Keeton, C.~R., Kochanek, C.~S., \& Seljak, U. 1997, \apj, 482, 604

\reference{} 
Keeton, C. R. et al. 2000, \apj, 542, 74

\reference{} 
Keeton, C.~R., Burles, S., Schechter, P., \& Wambsganss, J. 2005,
\apj in press


\reference{} 
Kochanek, C. S., Falco, E. E., Impey, C. D., Leh\'ar, J., McLeod, B. A., \&
Rix, H.-W. 1999, in After the Dark Ages: When Galaxies were Young (the 
Universe at 2 $\le z \le$ 5), ed. S. Holt \& E. Smith (New York: AIP), 163

\reference{} 
Kochanek, C.~S., Keeton, C.~R., \& McLeod, B.~A. 2001, \apj, 547, 50

\reference{} 
Kochanek, C. S., Schneider, P., Wambsganss, J., 2004, Part 2 of
Gravitational Lensing: Strong, Weak \& Micro, Proceedings of the 33rd Saas-Fee
Advanced Course, G. Meylan, P. Jetzer \& P. North, eds. (Springer-Verlag:
Berlin)

\reference{}
Kollmeier, J. et al. 2005, \apj\ submitted

\reference{}
Koopmans, L. V. E., Garrett, M. A., Blandford, R. D., Lawrence, C. R.,
Patnaik, A. R., \& Porcas, R. W., 2002, \mnras, 334, 39

\reference{}
Koopmans, L. V. E., Treu, T., Fassnacht, C. D., Blandford, R. D., \& Surpi, G.
2003, \apj, 599, 70

\reference{} 
Koopmans, L.~V.~E., \& Treu, T. 2004, Multiwavelength Cosmology.~Proceedings
of the "Multiwavelength Cosmology" conference, held on Mykonos Island, Greece,
17-20 June, 2003.~Ed. Manolis Plionis.~ASTROPHYSICS AND SPACE SCIENCE LIBRARY
Vol. 301.,~Pub. Kluwer Academic Publishers, Dordrecht, The Netherlands, 2004,
p.23, 23

\reference{} 
Kormendy, J. 2004, in Carnegie Observatories Astrophysics Series, Vol.  1:
Coevolution of Black Holes and Galaxies, ed. L. C. Ho (Cambridge: Cambridge
Univ. Press), in press.

\reference{}
Kormendy, J., \& Gebhardt, K. 2001, in 20th Texas Symposium on Relativistic 
Astrophysics, ed. H. Martel \& J.~C. Wheeler (Melville: AIP), 363

\reference{}
Kormendy, J., \& Richstone, D. 1995, \araa, 33, 581


\reference{} 
Kuhlbrodt, B., Orndahl, E., Wisotzki, L., \& Jahnke, K. 2005, astro-ph/0503284

\reference{}
Kukula, M., Dunlop, J.~S., McLure, R. J., Miller, L., Percival, W.~J.,
Baum, S. A., \& O'Dea, C.~P. 2001, \mnras, 326, 1533

\reference{}
Kundi\'c, T., Cohen, J.~G., Blandford, R.~D., \& Lubin, L.~M. 1997, \aj, 114, 507

\reference{}
Labb\'e, I., et al. 2005,  \apjl, 624, 81

\reference{}
Lacy, M., Gates, E. L., Ridgway, S. E., de Vries, W., Canalizo, G., Lloyd, J.
P., \& Graham, J. R. 2002, \aj, 124, 3023

\reference{}
Laor, A. 2001, \apj, 553, 677

\reference{}
Lawrence, C.~R., Elston, R., Januzzi, B.~T., \& Turner, E.~L. 1995, \aj, 110, 2570

\reference{}
Lawrence, A. 1999, Advances in Space Research, v. 23, 5-6, 1167

\reference{}
Leh\'ar, J., Falco, E. E., Kochanek, C.~S., McLeod, B.~A., Mu\~noz, J. A.,
Impey, C. D., Rix, H.-W., Keeton, C. R., \& Peng, C.~Y.  2000, \apj, 536, 854

\reference{} Lehnert, M.~D., Heckman, T.~M., Chambers, K.~C., \& Miley, G. K.
1992, \apj, 393, 68

\reference{} 
Lehnert, M. D., van Breugel, W. J. M., Heckman, T. M., \& Miley, G. K. 1999,
\apjs, 124, 11


\reference{} 
Lonsdale, C., Farrah, D., \& Smith, H. 2006, Review Article, ``Astrophysics
Update 2 -- topical and timely reviews on astronomy and astrophysics.'' Ed. J.
W. Mason, Springer/Praxis books. ISBN: 3-540-30312-X

\reference{}
MacAlpine, G.~M., \& Feldman, F.~R. 1982, \apj, 261, 412


\reference{}
Magorrian, J., et al.  1998, \aj, 115, 2285

\reference{} 
Marconi, A., \& Hunt, L. K. 2003, \apj, 589, L21


\reference{}
McLeod, K. K., \& Rieke G. H. 1994, \apjl, 420, 58

\reference{}
McLeod, K. K. 1995, \pasp, 107, 91

\reference{}
McLeod, K. K., \& Rieke G. H. 1995, \apjl, 454, L77

\reference{}
McLeod, K. K., \& McLeod, B. A. 2001, \apj, 546, 782

\reference{}
McLure, R. J., Kukula, M. J., Dunlop, J. S., Baum, S. A., O'Dea, C. P., \&
Hughes, D. H. 1999, \mnras, 308, 377

\reference{}
McLure, R. J., \& Dunlop, J. S. 2002, \mnras, 331, 795

\reference{}
McLure, R. J., \& Jarvis, M. J. 2002, \mnras, 337, 109

\reference{}
McLure, R. J., et al. 2005, astro-ph/0510121

\reference{}
Merloni, A., Rudnick, G., \& Di Matteo, T. 2004, \mnras, 354, 37

\reference{}
Merritt, D., \& Ferrarese, L. 2001, \mnras, 320, L30

\reference{}
Monier, E.~M., Turnshek, D.~A., \& Lupie, O.~L. 1998, \apj, 496, 177

\reference{}
Morgan, N.~D., Becker, R.~H., Gregg, M.~D., Schechter, P.~L., \& White, R.~L.
2001, \aj, 121, 611

\reference{}
Mu\~noz, J.~A., Falco, E.~E., Kochanek, C.~S., Leh\'ar, J., Herold, L.~K.,
Fletcher, A.~B., \& Burke, B.~F. 1998, \apjl, 492, 9


Mu\~noz, J.~A., et al. 2001, \apj, 546, 769

\reference{}
Mu\~noz, J. A., Kochanek, C. S. Keeton, C. R. 2001, \apj, 558, 657

\reference{} 
Mu\~noz, J. A., Falco, E. E., Kochanek, C.~S., McLeod, B.~A., Mediavilla, E.
2004, \apj, 605, 614

\reference{} 
Nelson, C.~H., 2000, \apj, 544, 91

\reference{}
Nelson, C.~H., Green, R. F., Bower, G., Gebhardt, K., \& Weistrop, D. 2004,
\apj, 615, 652

\reference{} 
Netzer, H. 2003, \apj, 583, L5

\reference{} 
Neugebauer, G., Matthews, K., Soifer, B.~T., \& Elias, J.~H.
1985, \apj, 298, 275

\reference{}
Oguri, M., et al. 2004, \pasj, 56, 399

\reference{} 
Onken, C. A. et al. 2004, \apj, 615, 645

\reference{}
Oscoz, A., Serra-Ricart, M., Medivilla, E., Buitrago, J., \& Goicoechea, L.~J. 1997, \apjl, 491, L7


\reference{}
Ouchi, M. et al. 2004, \apj, 611, 660



\reference{} 
Peng, C. Y., Ho, L, C., Impey, C. D., \& Rix, H.-W. 2002, \aj, 124, 266

\reference{} 
Peng, C. Y., Impey, C. D., Ho, L. C., Barton, E. J., \& Rix, H.-W.  2006,
\apj, 640, 114 [P06]

\reference{} 
Peterson, B. M. 1993, \pasp, 105, 247

\reference{} 
Peterson, B. M. et al. 2004, \apj, 613, 682

\reference{} 
Popovi\'c, L. C., et al. 2006, \apj, 637, 620

\reference{} 
Press, W.~H., Teukolsky, S.~A., Vetterling, W.~T., \& Flannery, B.~P. 1992, 
Numerical Recipes in Fortran (2nd Ed.; Cambridge: Cambridge Univ. Press) 

\reference{} 
Reddy, N. A., Steidel, C., \& Erb, D. 2005, in "Starbursts: From 30 Doradus to
Lyman Break Galaxies", eds. R. de Grijs \& R. M. Gonz\'alez Delgado,
Astrophysics \& Space Science Library, Vol. 329, Dordrecht: Springer, 205, p.
65

\reference{}
Reddy, N. A., et al. 2006, astro-ph/0602596

\reference{} 
Richards, G.~T., et al. 2004, \apj, 610, 679

\reference{} 
Ridgway, S.~E., Heckman, T.~M., Calzetti D., \& Lehnert, M. 2001, \apj, 550,
122

\reference{}
Rix, H.-W., de Zeeuw, P. T., Cretton, N., van der Marel, R. P., \& Carollo, C.
M. 1997, \apj, 488, 702



\reference{} 
Rudnick, G., et al. 2003, \apj, 599, 847

\reference{} 
Rusin, D. \& Kochanek, C.~S. 2005, \apj, 623, 666

\reference{}
S\'anchez, S. F., \& Gonz\'alez-Serrano, J. I. 2003, \aap, 406, 435

\reference{} 
S\'anchez, S.~F., et al. 2004, \apj, 614, 586

\reference{}
Schechter, P.,~L., Gregg, M.~D., Becker, R.~H., Helfand, D.~J., \& White, R.~L. 1998, \apj, 115, 1371

\reference{}
Schlegel, D.~J., Finkbeiner, D.~P., \& Davis, M. 1998, \apj, 500, 525

\reference{}
Schneider, D.~P., Lawrence, C.~R., Schmidt, M., Gunn, J.~E., Turner, E.~L., Burke, B.~F., \& Dhawan, V. 1985, \apj, 294, 66

\reference{} 
Schneider, P., Ehlers, J., \& Falco, E.~E.\ 1992, Gravitational Lenses, XIV,
560 pp.~112 figs..~Springer-Verlag Berlin Heidelberg New York.~ Also Astronomy
and Astrophysics Library.


\reference{} 
Shapley, A. E., et al. 2001, \apj, 562, 95

\reference{} 
Shapley, A. E., Steidel, C. C., Erb, D.~K., Reddy, N.~A., Adelberger, K.~L.,
Pettini, M., Barmby, P., \& Huang, J. 2005, \apj, 626, 698

\reference{}
Shields, G. A. et al. 2003, \apj, 583, 124

\reference{} 
Shields, G. A., Salviander, S., Bonning, E. W. 2006, astro-ph/0601675

\reference{}
Sluse, D., et al. 2003, \aap, 406, 43


\reference{}
Storrie-Lombardi, L.~J., McMahon, R.~G., Irwen, M.~J., \& Hazard, C. 1996, \apj, 468, 121

\reference{} 
Surpi, G., \& Blandford, R. D. 2003, \apj, 584, 100


\reference{} 
Treu, T., \& Koopmans, L. V. E. 2004, \apj, 611, 739

\reference{}
Treu, T., Malkan, M. A., \& Blandford, R. D. 2004, \apjl, 615, 97

\reference{} 
Treu, T., et al. 2005, astro-ph/0512044



\reference{}
van Dokkum, P. G., \& Franx, M. 2001, \apj, 553, 90

\reference{}
V\'eron-Cetty, M.-P., V\'eron, P. 1996, ESO Sci. Rep. No. 17. ESO Publications,
Garching


\reference{}
Vestergaard, M. 2002, \apj, 571, 733

\reference{}
Vestergaard, M. 2004, \apj, 601, 676

\reference{}
Vestergaard, M., \& Peterson, B. M. 2006, \apj, 641, 689

\reference{} 
Wandel, A., Peterson, B.~M., \& Malkan, M.~A. 1999, \apj, 526, 579


\reference{} 
Wang, Y.~P., Biermann, P.~L., \& Wandel, A. 2000, \aap, 361, 550

\reference{} 
Wayth, R. B., O'Dowd, M., \& Webster, R. L. 2005, \mnras, 359, 561

\reference{}
Winn, J.~N., et al. 2002, \aj, 123, 10

\reference{} 
Wiren, S. et al. 1992, \aj, 104, 1009

\reference{}
Wisotzki, L., K\"ohler, T., Ikonomou, M., \& Reimers, D. 1995, \aap, 297, L59

\reference{}
Wisotzki, L., K\"ohler, T., Lopez, S., \& Reimers, D. 1996, \aap, 315, L405

\reference{}
Wisotzki, L., Schechter, P.~L., Bradt, H.~V., Heinm\"uller, \& Reimers, D.
2002, \aap, 395, 17

\reference{}
Wisotzki, L, Schechter, P.~L., Chen, H.~W., Richstone, D., Jahnke, K.,
S\'anchez, S.~F., \& Reimers, D. 2004, \aap, 419, 31

\reference{} 
Woo, J.-H., \& Urry, C. M. 2002, \apj, 579, 530


\reference{}
Yoo, J., Kochanek, C. S., Falco, E. E., \& Mcleod, B. A. 2005,
astro-ph/0511001

\reference{}
Young, P., Sargent, W.~L.~W., Boksenberg, A., \& Oke, J.~B. 1981, \apj, 249, 415

\reference{}
Zhang, T.-J.\ 2004, \apjl, 602, L5 


\reference{} 
Zitelli, V., Mignoli, M., Zamorani, G., Marano, B., \& Boyle, B.  J. 1992,
\mnras, 256, 349

\end{references}

\clearpage


\begin{deluxetable}{|l|cc|cc|c|ccc|l|}
\tabletypesize{\scriptsize}
\tablewidth{0pt}
\tablecaption {Quasar and Host Galaxy Data from CASTLES}
\tablehead{
Object   &   $z_s$      & Size & RA   &  DEC  & $A_H$ &   Host   &  Quasar &  RLQ?  &  Comments \\
         &              & ($\arcsec$) & (J2000)  & (J2000) &  (mag)  &   (mag)  &  (mag)  &    &  \\
  (1)    &    (2)       &  (3) &     (4)     &        (5)      &  (6)  &    (7)   &   (8)   &   (9)   &  (10) }
\startdata
CTQ~414       &  1.29   & 1.22 & 01:58:41.44 & $-$43:25:04.20  &  0.01 &    19.87 &  18.09  &  ? & See Peng et al. (2005) \\
B~0712+472    &  1.34	& 1.46 & 07:16:03.58 &	+47:08:50.0    &  0.07 &    19.89 &  23.67  &  Y & Lens edge on disk?\\
SBS~0909+532  &	 1.38	& 1.17 & 09:13:01.05 &	+52:59:28.83   &  0.01 &    20.89 &  16.41  &  ? & \\
FBQ~0951+2635 &	 1.24	& 1.11 & 09:51:22.57 &	+26:35:14.1    &  0.01 &    21.07 &  16.57  &  ? & Lens edge on disk\\
Q~0957+561    &  1.41   & 6.26 & 10:01:20.78 &  +55:53:49.4    &  0.01 &    17.83 &  17.05  &  Y & \\
LBQS~1009$-$0252c& 1.63 & ---  & 10:12:15.83 & $-$03:07:01.6   &  0.02 &    19.41 &  17.90  &  ? & Not lensed \\
B~1030+071    &  1.54	& 1.56 & 10:33:34.08 &	+07:11:25.5    &  0.01 &    19.93 &  19.15  &  Y & \\
RXJ~1131$-$1231 & 0.66  & 3.69 &  11:31:51.6 &  $-$12:31:57    &  0.02 &    17.48 &  19.25  &  ? & B/D decomposition done \\
SDSS~1226$-$0006& 1.12  & 1.20 & 12:26:08.10 &  $-$00:06:02.0  &  0.01 &    19.77 &  18.93  &  ? & \\
SDSS~1335+0118&  1.57   & 1.57 & 13:35:34.79 &  +01:18:05.5    &  0.01 &    20.24 &  17.49  &  ? & \\
B~1600+434    &  1.59	& 1.40 & 16:01:40.45 &	+43:16:47.8    &  0.01 &    20.23 &  21.43  &  Y & Lens edge on disk \\
FBQ~1633+3134 &  1.52   & 0.75 & 16:33:48.99 &	+31:34:11.90   &  0.02 &    19.84 &  16.84  &  ? & \\
B~2045+265    &  1.28	& 2.76 & 20:47:20.35 &	+26:44:01.2    &  0.13 &    22.81 &  24.98  &  Y & \\
MGC~2214+3550 A&  0.88   & ---  & 22:14:57.04 &  +35:51:25.5    &  0.07 &    18.03 &  19.16  &  Y & Not lensed \\
MGC~2214+3550 B&  0.88   & ---  & 22:14:57.04 &  +35:51:25.5    &  0.07 &    19.16 &  20.08  &  Y & Not lensed \\
\tableline
HE~0047$-$1756 & 1.66   & 1.44 & 00:50:27.83 &  $-$17:40:8.8   &  0.01 &    19.72 &  18.25  &  ? & \\
Q~0142$-$110  &	2.72	& 2.24 & 00:49:41.89 &	$-$27:52:25.7  &  0.02 &    21.01 &  16.55  &  ? & \\
MG~0414+0534  &	2.64	& 2.40 & 04:14:37.73 &	+05:34:44.3    &  0.18 &    20.78 &  18.63  &  Y & \\
HE~0435$-$1223 &  1.69  & 2.43 & 04:38:14.9  &	$-$12:17:14.4  &  0.03 &    20.09 &  19.64  &  ? & \\
RXJ~0911+0551 & 2.80	& 2.47 & 09:11:27.50 &	+05:50:52.0    &  0.03 &    22.23 &  20.23  &  ? & \\
RXJ~0921+4528 & 1.65	& ---  & 09:21:12.81 &	+45:29:04.4    &  0.01 &    20.01 &  16.95  &  ? & Not lensed? \\
RXJ~0921+4528 & 1.65	& ---  & 09:21:12.81 &	+45:29:04.4    &  0.01 &    19.34 &  18.36  &  ? & Not lensed? \\
SDSS~0924+0219 & 1.54   & 1.74 & 09:24:55.87 &  +02:19:24.9    &  0.03 &    19.85 &  21.00  &  ? & \\
BRI~0952$-$0115 & 4.5   & 1.00 & 09:55:00.01 &	$-$01:30:05.0  &  0.01 &    22.29 &  19.10  &  ? & \\
J~1004+1229   &  2.65   & 1.54 & 10:04:24.9  &  +12:29:22.3    &  0.02 &    20.08 &  17.28  &  ? & \\
LBQS~1009$-$0252 & 2.74 & 1.54 & 10:12:15.71 &	$-$03:07:02.0  &  0.02 &    20.75 &  17.75  &  ? & \\
Q~1017$-$207  &  2.55	& 0.85 & 10:17:24.13 &	$-$20:47:00.4  &  0.03 &    19.94 &  17.29  &  ? & \\
HE~1104$-$1805   & 2.32 & 3.19 & 11:06:33.45 &	$-$18:21:24.2  &  0.03 &    20.14 &  18.27  &  ? & \\
PG~1115+080   &  1.72	& 2.32 & 11:18:17.00 &	+07:45:57.7    &  0.02 &    20.23 &  18.80  &  ? & \\
H~1413+117    &  2.55	& 1.35 & 14:15:46.40 &	+11:29:41.4    &  0.01 &    20.47 &  18.41  &  ? & \\
B~1422+231    &  3.62	& 1.68 & 14:24:38.09 &	+22:56:00.6    &  0.03 &    [21]  &  16.71  &  Y & Upper limit on host \\
SBS~1520+530  &  1.86	& 1.59 & 15:21:44.83 &	+52:54:48.6    &  0.01 &    21.09 &  19.01  &  ? & Lens edge on disk \\
PMNJ~1632$-$0033 & 3.42 & 1.47 & 16:32:57.68 &  $-$00:33:21.1  &  0.06 &    22.25 &  19.59  &  ? & \\
MG~2016+112   &  3.27   & 3.52 & 20:19:18.15 &	+11:27:08.3    &  0.14 &    22.35 &  21.47  &  Y & NL QSO \\
HE~2149$-$2745  &  2.03 & 1.70 & 21:52:07.44 &	$-$27:31:50.2  &  0.02 &    19.75 &  16.85  &  ? & \\
PSS~2322+1944   &  4.1  & 1.50 & 23:22:07.2  &	+19:44:23      &  0.03 &    21.83 &  18.99  &  ? & \\
\tableline
\enddata 
\tablecomments {
Col. (1): Object name.
Col. (2): Quasar redshift.
Col. (3): Estimated angular diameter of lensing geometry.
Col. (4): RA (J2000).
Col. (5): DEC (J2000).
Col. (6): Extinction in $H$-band.
Col. (7): Intrinsic (i.e. lens deprojected) $H$-band host magnitude.  The typical
	  uncertainties are 0.3 mag.
Col. (8): Intrinsic $H$-band quasar magnitude.
Col. (9): Radio loud quasar? Objects with ``?'' do not have known radio observations.
Col. (10): Comments.
}
\end{deluxetable}


\begin{deluxetable}{|ll|rcccc|ccc|l|}
\tabletypesize{\scriptsize}
\tablewidth{0pt}
\tablecaption {Quasar and Host Galaxy Data Compiled from the Literature}
\tablehead{Object & $z$ & Filter & \multicolumn{4}{c|} {NICMOS Mag} & \multicolumn{2}{|c}{Quasar} & Radio  & References/Comments\\
     &          &        & Host  &  Err  &  Quasar  &  Err &    Filter  & Mag  & Loud? & \\
    (1)         &   (2)  &  (3)   & (4)    & (5) &  (6)  & (7) & (8)   & (9) & (10) & (11)}
\startdata
SGP5:46     	& 0.955  & F110M  & 20.09  & 0.4 & 19.45 & 0.3 & $V^*$ & 19.7 & N  & 1, assumed quasar ($B-V$)=0.3 \\
PKS~0440$-$00  	& 0.844  & F110M  & 18.79  & 0.4 & 18.42 & 0.3 & $V^*$ & 19.1 & Y  & 1 \\
PKS~0938+18	& 0.943  & F110M  & 19.46  & 0.4 & 19.78 & 0.3 & $V$ & 18.9 & Y  & 1, assumed quasar ($B-V$)=0.3 \\
3C~422		& 0.942  & F110M  & 18.24  & 0.4 & 17.85 & 0.3 & $V$ & 18.9 & Y  & 1, assumed quasar ($B-V$)=0.3 \\
\tableline
SGP2:11		& 1.976  & F165M  & 20.64  & 0.75 & 18.96 & 0.3 & $B^*$ & 20.9 & N  & 1 \\
SGP2:25		& 1.868  & F165M  & 19.88  & 0.75 & 19.59 & 0.3 & $B^*$ & 20.7 & N  & 1 \\
SGP2:36		& 1.756  & F165M  & 19.73  & 0.75 & 19.97 & 0.3 & $B^*$ & 20.7 & N  & 1 \\
SGP3:39		& 1.964  & F165M  & 19.75  & 0.75 & 19.53 & 0.3 & $B^*$ & 20.8 & N  & 1 \\
SGP4:39		& 1.716  & F165M  & 21.59  & 0.75 & 18.85 & 0.3 & $B^*$ & 20.8 & N  & 1 \\
4C~45.51	& 1.992  & F165M  & 17.79  & 0.75 & 17.41 & 0.3 & $B$ & 20.1 & Y  & 1 \\
\tableline
MZZ~9744    	& 2.735  & F160W  & 21.73  & 0.5 & 20.02 & 0.05 & $B^*$ & 21.4 & N & 2 \\
MZZ~9592    	& 2.710  & F160W  & 20.70  & 0.1 & 19.57 & 0.03 & $B^*$ & 21.8 & N & 2 \\
MZZ~1558    	& 1.829  & F165M  & 20.64  & 0.2 & 19.08 & 0.04 & $B^*$ & 21.5 & N & 2 \\
MZZ~11408   	& 1.735  & F165M  & 20.78  & 0.4 & 21.08 & 0.06 & $B^*$ & 21.9 & N & 2 \\
MZZ~4935    	& 1.876  & F165M  & 22.00  & 0.4 & 21.23 & 0.06 & $B^*$ & 21.8 & N & 2 \\
\tableline
\enddata 
\tablecomments {
Col. (1): Object name.
Col. (2): Redshift.
Col. (3): {\it HST} Filter.
Col. (4-7): Apparent magnitude and their published uncertainties, in the Vega
	    magnitude system, corrected for extinction from Schlegel et al.
	    (1998).  The values are taken from Kukula et al. (2001),
	    Ridgway et al. (2001), and corrected slightly by P06 for morphology
	    assumptions.
Col. (8/9): Quasar magnitude (corrected for extinction), in the Vega magnitude
	    system, corresponding to the filter in Col. (8).  Filters with
	    superscript $^*$ are photographic magnitudes.  The $B$ and
	    $V$-band magnitudes of the quasars in Kukula et al. (2001) sample
	    are from V\'eron-Cetty \& V\'eron (1996), and references therein,
	    while the $B$-band magnitudes for MZZ objects are from Zitelli et
	    al.  (1992).  Where $V$-band magnitude is needed and unavailable,
	    we used $(B-V)=0.3$, which corresponds to $f_\nu\propto\nu^{-0.5}$.
Col. (10): Radio-loud quasar or radio-quiet quasar.
Col. (11): The photometry for each set of objects comes from the references
          shown.  
References.--- (1) Kukula et al. 2001; (2) Ridgway et al. 2001.
}
\end{deluxetable}

\clearpage
\thispagestyle{empty}
\begin{landscape}
\begin{deluxetable}{|l|cc|c|ccccc|c|l|}
\tabletypesize{\scriptsize}
\tablewidth{0pt}
\tablecaption {Quasar and Host Galaxy Derived Quantities From CASTLES Data}
\tablehead{
Object   & $z_s$ & Diameter    &    Host     &   \multicolumn{5}{|c|}{Quasar Related Quantities} &       \mbh        &  Comments/References \\
         &       &             &    $M_R$    &  $M_B$   &   Line(s)  &  FWHM & log ($\lambda$L$_\lambda$(cont)) & Spectral slope &                & \\
         &       & [$\arcsec$] &   [mag]     &  [mag]   &   used          & [\AA] &  log ([ergs s$^{-1}$]) &  $n$  & [$10^9$ \msol] & \\
  (1)    &    (2)       &  (3) &   (4)       &    (5)   &       (6)              &   (7)   &   (8)  & (9)  & (10)  &  (11) }
\startdata
CTQ~414      &   1.29   & 1.22 & $-$22.77    & $-$24.19  & C~{\sc iv}/Mg~{\sc ii} & 40/40   & 45.53/45.31 & $-1.65$ & 0.34/0.16 & See Peng et al. (2005)\\
B~0712+472   &   1.34	& 1.46 & $-$22.86    & $-$17.90  & Mg~{\sc ii}            & 110     & 42.74 &   0           & 0.07 & 1, Lens edge on disk?\\
SBS~0909+532 & 	 1.38	& 1.17 & $-$21.95    & $-$25.70  & Mg~{\sc ii}            & 138     & 45.95 & $-$0.95       & 3.87 & 2, Confusion lim./PSF issues \\
SDSS~0924+0219 & 1.54   & 1.74 & $-$23.32    & $-$21.78	 & Mg~{\sc ii}		  & 61	    & 44.35 & $-$1.9	    & 0.11 & 3 \\
FBQ~0951+2635 &	 1.24	& 1.11 & $-$21.22    & $-$25.66  & Mg~{\sc ii}            & 70      & 45.90 & $-$1.73       & 0.89 & 4, Lens edge on disk\\
Q~0957+561   &   1.41   & 6.26 & $-$25.08    & $-$27.00  & C~{\sc iv}/Mg~{\sc ii} & 50/100  & 46.71/46.43 & $-1.8$  & 2.01/3.02 & 5\\
LBQS~1009$-$0252c & 1.63 & --- & $-$23.94    & $-$25.07  & C~{\sc iv}             & 71      & 46.11 & $-$2.1        & 1.64 & 6, Not lensed \\
B~1030+071   &   1.54	& 1.56 & $-$23.25    & $-$23.27  & Mg~{\sc ii}            & 86      & 44.82 & $-$1.0        & 0.35 & 1 \\
RXJ~1131$-$1231 & 0.66  & 3.69 & $-$23.14    & $-$20.90  &  H$\beta$              & 90      & 43.95 & $-$1.0        & 0.06 & 7 \\
SDSS~1226$-$0006& 1.12  & 1.20 & $-$22.44    & $-$22.70  & Mg~{\sc ii}            & 120     & 44.48 & $-$0.6        & 0.68 & \\
SDSS~1335+0118 & 1.57   & 1.57 & $-$23.03    & $-$25.17	 & Mg~{\sc ii}		  & 120	    & 45.60 & $-$1.0	    & 1.55 & 8 \\
B~1600+434   &   1.59	& 1.40 & $-$23.04    & $-$21.41  & Mg~{\sc ii}            & 63      & 44.23 & $-$1.9        & 0.10 & 1, Lens edge on disk \\
FBQ~1633+3134 &  1.52   & 0.75 & $-$23.30    & $-$25.85  & C~{\sc iv}/Mg~{\sc ii} & 65/104  & 46.24/45.98 & $-$1.75 & 1.76/1.84 & 9 \\
B~2045+265   &   1.28	& 2.76 & $-$19.80    & $-$15.91  & Mg~{\sc ii}            & 78      & 41.49 &  1.3          & 0.01 & 10 \\
MGC~2214+3550 A& 0.88   & ---  & $-$23.42    & $-$21.82	 & Mg~{\sc ii}		  & 160	    & 44.46 & [$-$1.5]      & 1.64 & 11, Not lensed\\
MGC~2214+3550 B& 0.88   & ---  & $-$22.08    & $-$20.97	 & Mg~{\sc ii}		  & 110	    & 44.08 & [$-$1.5]	    & 0.43 & 11, Not lensed\\
\tableline
HE~0047$-$1756&  1.66   & 1.44 & $-$23.69    & $-$24.75	 & C~{\sc iv}		  & 70	    & 46.07 & $-2.1$	    & 1.48 & 12 \\
Q~0142$-$110  &  2.72	& 2.24 & $-$24.02    & $-$27.37  & C~{\sc iv}             & 75      & 46.85 & $-$1.75       & 2.26 & 13, line center absorbed \\
MG~0414+0534  &  2.64	& 2.40 & $-$24.13    & $-$25.22  & H$\beta$               & 262     & 45.73 &   1           & 1.82 & 14 \\
HE~0435$-$1223 & 1.69   & 2.43 & $-$23.40    & $-$23.29  & C~{\sc iv}             & 70      & 45.19 & $-$1.7        & 0.50 & 15 \\
RXJ~0911+0551 &  2.80	& 2.47 & $-$22.93    & $-$23.73  & C~{\sc iv}             & 100     & 45.57 & $-$2.07       & 0.80 & 16 \\
RXJ~0921+4528A &  1.65	& 6.97 & $-$23.39    & $-$25.95  & C~{\sc iv}/Mg~{\sc ii} & 52/141  & 46.31/46.03 & $-$1.8  & 1.10/3.21 & 17, Excl. from fit: not lensed? \\
RXJ~0921+4528B &  1.65	& 6.97 & $-$24.06    & $-$24.54  & C~{\sc iv}/Mg~{\sc ii} & 52/141  & 45.74/45.46 & $-$1.8  & 0.55/1.74 & 17, Excl. from fit: not lensed? \\
BRI~0952$-$0115  & 4.5  & 1.00 & $-$25.47    & $-$26.09  & C~{\sc iv}             & 146     & 46.00 & $-$1.1        & 1.39 & 18 \\
J~1004+1229   &  2.65   & 2.65 & $-$24.86    & $-$26.58	 & C~{\sc iv}		  & 91	    & 46.41 & $-$1.5	    & 2.02 & 19 \\
LBQS~1009$-$0252 & 2.74 & 1.54 & $-$24.31    & $-$26.19  & C~{\sc iv}/Mg~{\sc ii} & 75/100  & 46.27/46.08 & $-$1.55 & 1.11/0.86 & 6 \\
Q~1017$-$207  &  2.55	& 0.85 & $-$24.83    & $-$26.49  & C~{\sc iv}             & 80      & 46.42 & $-$1.6        & 1.68 & 20 \\
HE~1104$-$1805 & 2.32   & 3.19 & $-$24.31    & $-$25.35  & C~{\sc iv}             & 103     & 46.18 & $-$2.02       & 2.37 & 21 \\
PG~1115+080   &  1.72	& 2.32 & $-$23.29    & $-$24.24  & C~{\sc iv}/Mg~{\sc ii} & 69/127  & 45.74/45.39 & $-$2.02 & 0.92/1.23 & 22 \\
H~1413+117    &  2.55	& 1.35 & $-$24.30    & $-$25.35  & C~{\sc iv}             & 52      & 45.61 & $-$0.9        & 0.26 & 23, BALQSO \\
B~1422+231    &  3.62	& 1.68 & [$-$24.6]   & $-$27.90  & C~{\sc iv}/Mg~{\sc ii} & 133/139 & 46.88/46.74 & $-$1.4  & 4.79/2.23 & 24, Excl. from fit: non-det.\\
SBS~1520+530  &  1.86	& 1.59 & $-$22.66    & $-$24.16  & C~{\sc iv}             & 75      & 45.64 & $-$1.9        & 0.88 & 25, Lens edge on disk \\
PMNJ~1632$-$0033& 3.42  & 1.47 & $-$23.97    & $-$24.84  & C~{\sc iv} 		  & 100	    & 45.22 & $-$0.7	    & 0.39 & 26 \\
MG~2016+112   &  3.27   & 3.52 & $-$23.59    & $-$23.00  & C~{\sc iv}             & 25      & 44.56 & $-$0.7        & 0.01 & 27, Excl. from fit: NLQSO \\
HE~2149$-$2745  &  2.03 & 1.70 & $-$24.27    & $-$26.52  & C~{\sc iv}             & 117     & 46.67 & $-$2.05       & 6.62 & 28, BALQSO \\
PSS~2322+1944   &  4.1  & 1.50 & $-$25.48    & $-$25.8  & $\epsilon_{\mbox{Edd}}$=1      & N/A     &  N/A  &  N/A   & $\gtrsim 2.36$ & 29 \\
\tableline
\enddata 
\tablecomments {
Values in square brackets [] indicate that they are held fixed.
Col. (1): Object name.
Col. (2): Quasar redshift.
Col. (3): Estimated angular diameter of lensing geometry.
Col. (4): Inferred host restframe absolute $R$-band magnitude based on Sbc SED.
Col. (5): Inferred quasar restframe $B$-band magnitude.
Col. (6): Emission line(s) used to infer \mbh.
Col. (7): Observed-frame FWHM of emission line in Col. 6.
Col. (8): Monochromatic continuum luminosity appropriate for corresponding
          lines used in Col. 6: for C~{\sc iv}, $\lambda$L$_\lambda(1350\mbox{ \AA})$;
	  for Mg~{\sc ii}, $\lambda$L$_\lambda(3000\mbox{ \AA})$;
	  for H$\beta$, $\lambda$L$_\lambda(5100\mbox{ \AA})$.
Col. (9): Continuum spectral slope $n$ in $\lambda^n$ used to extrapolate 
	  values in Cols. 5 and 8.
Col. (10): Black hole masses using the virial technique, using the
           corresponding lines in Col. 6.  For PSS~2322+1944, \mbh\ is
	   based on Eddington limited accretion in Isaak et al. (2002),
	   corrected for lensing magnification.  Thus it is a lower limit.
Col. (11): Comments and References from which spectral information was obtained ---
    1. Fassnacht \& Cohen (1998),
    2. Oscoz et al. (1997),
    3. Inada et al. (2003),
    4. Schechter et al. (1998),
    5. Young et al. (1981),
    6. Hewett et al. (1994),
    7. Sluse et al. (2003),
    8. Oguri et al. (2004),
    9. Morgan et al. (2001),
   10. Fassnacht et al. (1999),
   11. Mu\~noz et al. (1998),
   12. Wisotzki et al. (2004),
   14. MacAlpine \& Feldman (1982),
   14. Lawrence et al. (1995),
   15. Wisotzki et al. (2002),
   16. Bade et al. (1997),
   17. Mu\~noz et al. (2001),
   18. Storrie-Lombardi et al. (1996),
   19. G\'omez-Alvarez (2005),
   20. Claeskens, Surdej, \& Remy (1996),
   21. Wisotzki et al. (1995),
   22. Kundi\'c et al. (1997),
   23. Monier, Turnshek, \& Lupie (1998),
   24. Hammer et al. (1995),
   25. Chavushyan et al. (1997),
   26. Winn, J.~N., et al. (2002),
   27. Schneider et al. (1985),
   28. Wisotzki et al. (1996),
   29. Cox, P., et al. (2002),
}
\end{deluxetable}
\clearpage
\end{landscape}


\begin{deluxetable}{|l|c|c|ccccc|c|l|}
\tabletypesize{\scriptsize}
\tablewidth{0pt}
\tablecaption {Quasar and Host Galaxy Derived Quantities From the Literature}
\tablehead{
Object   & $z_s$ &     Host      &   \multicolumn{5}{|c|}{Quasar Related Quantities} &       \mbh    &  Comments  \\
         &       &    $M_R$      &  $M_B$   &   Line(s)   &  FWHM & log ($\lambda$L$_\lambda$(cont)) & Spectral slope  &       &    \\
         &       &    [mag]      &  [mag]   &   used      & [\AA] &  log ([ergs s$^{-1}$]) &   $n$     & [$10^9$ \msol] &        \\
  (1)    &       (2)  &  (3)     &   (4)    &    (5)      & (6) &  (7)  & (8)  & (9) & (10)}
\startdata
SGP5:46     &   0.955 & $-$22.90 & $-$23.04 & Mg~{\sc ii} &  55 & 44.96 & $-$2.2 & 0.16 & Narrow line Mg~{\sc ii}\\
PKS~0440$-$00 & 0.844 & $-$23.83 & $-$23.69 & Mg~{\sc ii} &  50 & 45.16 & $-$1.9 & 0.18 & Narrow line Mg~{\sc ii}\\
PKS~0938+18 &   0.943 & $-$23.49 & $-$22.76 & Mg~{\sc ii} & 100 & 44.90 & $-$2.5 & 0.49 & \\
3C~422      &   0.942 & $-$24.71 & $-$24.39 & Mg~{\sc ii} & 140 & 45.36 & $-$1.4 & 1.58 & Heavily absorbed Mg~{\sc ii}\\
\tableline			   	  	   	   			  
SGP2:11     &   1.976 & $-$23.19 & $-$24.20 & C~{\sc iv}  &  90 & 45.57 & $-$1.7 & 1.06 & \\
SGP2:25     &   1.868 & $-$23.77 & $-$23.56 & C~{\sc iv}  &  70 & 45.48 & $-$2.0 & 0.61 & \\
SGP2:36     &   1.756 & $-$23.75 & $-$23.22 & C~{\sc iv}  & 100 & 45.61 & $-$2.5 & 1.58 & \\
SGP3:39     &   1.964 & $-$24.05 & $-$23.76 & C~{\sc iv}  &  85 & 45.66 & $-$2.2 & 1.06 & \\
SGP4:39     &   1.716 & $-$21.82 & $-$24.00 & C~{\sc iv}  &  45 & 45.38 & $-$1.5 & 0.25 & \\
4C~45.51    &   1.992 & $-$26.06 & $-$25.66 & C~{\sc iv}/Mg~{\sc ii} & 58/101 & 45.90/45.83 & $-$1.2  & 0.65/0.58 & \\
\tableline			   	  	   	   			  
MZZ~9744    &   2.735 & $-$23.33 & $-$23.91 & C~{\sc iv}  &  90 & 45.62 & $-$2.0  & 0.71 & \\
MZZ~9592    &   2.71  & $-$24.32 & $-$24.35 & C~{\sc iv}  &  90 & 45.52 & $-$1.5  & 0.63 & C~{\sc iv} partly absorbed\\
MZZ~1558    &   1.829 & $-$22.95 & $-$23.82 & C~{\sc iv}  & 120 & 45.22 & $-$1.3  & 1.34 & \\
MZZ~11408   &   1.735 & $-$22.66 & $-$22.07 & C~{\sc iv}  &  50 & 45.09 & $-$2.4  & 0.21 & \\
MZZ~4935    &   1.876 & $-$21.67 & $-$22.11 & C~{\sc iv}  &  50 & 45.21 & $-$2.6  & 0.22 & \\
\tableline
\enddata
\tablecomments {
Col. (1): Object name.
Col. (2): Quasar redshift.
Col. (3): Inferred host restframe absolute $R$-band magnitude based on Sbc SED.
Col. (4): Inferred quasar restframe $B$-band magnitude.
Col. (5): Emission line(s) used to infer \mbh.
Col. (6): Observed-frame FWHM of emission line in Col. 5.
Col. (7): Monochromatic continuum luminosity appropriate for corresponding
          lines used in Col. 6: for C~{\sc iv}, $\lambda$L$_\lambda(1350\mbox{ \AA})$;
          for Mg~{\sc ii}, $\lambda$L$_\lambda(3000\mbox{ \AA})$.
Col. (8): Continuum spectral slope $n$, in $\lambda^n$, used to extrapolate 
          values in Cols. 4 and 7.
Col. (9): Black hole masses using the virial technique, using the
           corresponding lines in Col. 5. 
}
\end{deluxetable}

\clearpage

\end {document}